\documentclass[camera,letterpaper,nomarginnotes,nonarrowgutter,hyphens]{jpaper}

\usepackage[colorlinks]{hyperref}
\usepackage[dvipsnames]{xcolor}
\usepackage{soul}
\usepackage{xspace}
\usepackage{multirow}
\usepackage{tabularx}
\usepackage{tikz}
\usepackage[bottom]{footmisc}
\newcommand*\circled[1]{\tikz[baseline=(char.base)]{\node[shape=circle,fill,inner sep=0.5pt] (char) {\textcolor{white}{#1}};}}
\usepackage[most]{tcolorbox}
\usepackage{amsmath}
\usepackage{enumitem}

\usepackage{subcaption}
\usepackage{fancyhdr}
\usepackage{mathtools}

\usepackage{booktabs}

\usepackage{soul}

\definecolor{darkspringgreen}{rgb}{0.09, 0.45, 0.27}
\definecolor{denim}{rgb}{0.08, 0.38, 0.74}
\definecolor{darkolivegreen}{rgb}{0.33, 0.42, 0.18}
\definecolor{tangerine}{rgb}{0.95, 0.52, 0.0}
\definecolor{mahogany}{rgb}{0.75, 0.25, 0.0}
\definecolor{coolblack}{rgb}{0.0, 0.18, 0.39}
\definecolor{darkpink}{rgb}{0.88, 0.28, 0.54}

\definecolor{seagreen}{rgb}{0.18, 0.55, 0.34}

\usepackage[colorinlistoftodos,prependcaption,textsize=scriptsize,textwidth=13mm]{todonotes} %
\usepackage{marginnote} 
\let\marginpar\marginnote
\newcommand\revmid[1]{\todo[linecolor=magenta,backgroundcolor=magenta!15,bordercolor=magenta]{\textcolor{black}{\textbf{#1}}}}

\renewcommand\revmid[1]{}

\newcommand{\js}[1]{{\color{black}{#1}}} %

\definecolor{pred}{rgb}{0.7843, 0.0039, 0.3137} %

\definecolor{darkpink}{rgb}{0.88, 0.28, 0.54}
\definecolor{forestgreen}{rgb}{0.0, 0.27, 0.13}
\definecolor{amber}{rgb}{1.0, 0.49, 0.0}

\newcommand{\joel}[1]{{\color{cyan}#1}}

\newcommand{\inum}[1]{(\textit{#1})\xspace}

\newcommand{\sect}[1]{{§#1}\xspace} %
\newcommand{\head}[1]{{\noindent\textbf{#1.}\xspace}} %
\newcommand{\figs}[1]{{Figs.~#1}\xspace} %
\newcommand{\fig}[1]{{Fig.~#1}\xspace} %

\newcommand\proposal{MegIS\xspace}
\newcommand\proposals{MegIS's\xspace}

\newcommand\dram{\texttt{\omciv{No-I/O}}\xspace}

\newcolumntype{Y}{>{\centering\arraybackslash}X}
\usepackage{tikz}

\newcommand{\squishlist}{
 \begin{list}{$\circ$}
  { \setlength{\itemsep}{0pt}
     \setlength{\parsep}{0pt}
     \setlength{\topsep}{3pt}
     \setlength{\partopsep}{0pt}
     \setlength{\leftmargin}{1em}
     \setlength{\labelwidth}{1em}
     \setlength{\labelsep}{0.5em} } }

\newcommand{\squishend}{
  \end{list}  }

\usepackage[compact]{titlesec}
\titlespacing\section{0pt}{4pt plus 2pt minus 2pt}{4pt plus 2pt minus 2pt}
\titlespacing\subsection{0pt}{4pt plus 2pt minus 2pt}{4pt plus 2pt minus 2pt}
\titlespacing\subsubsection{0pt}{4pt plus 2pt minus 2pt}{4pt plus 2pt minus 2pt}
\makeatletter
\g@addto@macro{\normalsize}{%
  \setlength{\abovedisplayskip}{5pt plus 0.5pt minus 1pt}
  \setlength{\belowdisplayskip}{5pt plus 0.5pt minus 1pt}
  \setlength{\abovedisplayshortskip}{3pt}
  \setlength{\belowdisplayshortskip}{3pt}
  \setlength{\intextsep}{3pt plus 1pt minus 1pt}
  \setlength{\textfloatsep}{3pt plus 1pt minus 1pt}
  \setlength{\skip\footins}{3pt plus 1pt minus 1pt}}
\makeatother

\DeclareRobustCommand\wcirc[1]{\tikz[baseline=(char.base)]{           \node[shape=circle,draw,inner sep=0pt,fill=white, text=black] (char) {#1};}}

 \usepackage{setspace}
\usepackage[colorinlistoftodos,prependcaption,textsize=scriptsize]{todonotes}
\usepackage{todonotes}

\usepackage{blindtext,graphicx}
\usepackage[absolute]{textpos}
\usepackage{datetime}

\definecolor{seagreen}{rgb}{0.18, 0.55, 0.34}
\definecolor{ballblue}{rgb}{0.13, 0.67, 0.8}

\newcommand\omciv[1]{{{#1}}}

\newcommand\omcviii[1]{{{#1}}}

\definecolor{darkgreen}{rgb}{0.0, 0.44, 0.34}
 
 \newcommand\ssdc{\texttt{SSD-C}\xspace}
 \newcommand\ssdp{\texttt{SSD-P}\xspace}
 
\sloppypar

\newcommand\hm[1]{{\color{violet}{#1}}}

\newcommand\randomio{\textsf{R-Qry}\xspace}
\newcommand\streamio{\textsf{S-Qry}\xspace}

\definecolor{dollarbill}{rgb}{0.52, 0.73, 0.4}

\newcommand\cmashopt{\textsf{KSS}\xspace}

\renewcommand{\js}[1]{{\color{black}{#1}}} %
\renewcommand{\joel}[1]{{\color{black}#1}}
\renewcommand\hm[1]{{\color{black}{#1}}}

\definecolor{indiagreen}{rgb}{0.07, 0.53, 0.03}

\newcommand\rev[1]{{\color{black}{#1}}} %
\newcommand{\new}[1]{{\color{black}{#1}}}  %

\newcommand\revhid[1]{\todo[linecolor=magenta,backgroundcolor=magenta!15,bordercolor=magenta]{\textcolor{black}{\textbf{#1}}}}

\newcommand\revh[1]{{\color{black}{#1}}} %
\newcommand\gram[1]{{\color{black}{#1}}}

\renewcommand\revhid[1]{}

\newcommand\omf[1]{{\color{black}{#1}}}

\newcommand\hhl[1]{{\color{cornflowerblue}{#1}}}
\newcommand\marklabel[1]{{\noindent\hyperref[rev:#1]{\markqg{#1}}}}

\hypersetup{
citecolor=blue,
linkcolor=blue,
urlcolor=blue
}

\definecolor{lightblue}{rgb}{0.68, 0.85, 0.9}

\DeclareRobustCommand{\markqg}[1]{{\noindent\textbf{{\sethlcolor{cornflowerblue!40}{\hl{#1}}}}}}

\definecolor{cornflowerblue}{rgb}{0.39, 0.58, 0.93}
\definecolor{brandeisblue}{rgb}{0.0, 0.44, 1.0}

\newcommand\irev[1]{{\color{cornflowerblue}{#1}}} %

\newcommand\revid[1]{\todo[linecolor=cornflowerblue,backgroundcolor=cornflowerblue!50,bordercolor=cornflowerblue]{\textcolor{black}{\textbf{#1}}}}

\fancypagestyle{firstpage}
{

    \fancyfoot[C]{\thepage}
}

\renewcommand\revid[1]{}
\renewcommand\irev[1]{{\color{black}{#1}}} 
\renewcommand\hhl[1]{{\color{black}{#1}}}

\definecolor{darkspringgreen}{rgb}{0.09, 0.45, 0.27}
\definecolor{lasallegreen}{rgb}{0.03, 0.47, 0.19}
\definecolor{seagreen}{rgb}{0.18, 0.55, 0.20}
\definecolor{coral}{rgb}{1.0, 0.5, 0.31}

\newcommand\bback[1]{{\color{black}{#1}}} %

\newcommand\icut[1]{} %
\definecolor{gray(x11gray)}{rgb}{0.65, 0.65, 0.65}

\newcommand\omi[1]{{\color{magenta}{#1}}} %

\newcommand\mganalysis{metagenomic analysis\xspace}

\newcommand\omn[1]{\todo[linecolor=magenta,backgroundcolor=magenta!15,bordercolor=magenta]{\textcolor{black}{#1}}} %
\definecolor{deepcarrotorange}{rgb}{0.91, 0.41, 0.17}
\definecolor{forestgreen(web)}{rgb}{0.13, 0.55, 0.13}
\definecolor{burntsienna}{rgb}{0.91, 0.45, 0.32}
\definecolor{cadmiumorange}{rgb}{0.93, 0.53, 0.18}
\newcommand\omii[1]{{\color{orange}{#1}}} %
\newcommand\omiii[1]{{\color{black}{#1}}} %

\renewcommand\omi[1]{{\color{black}{#1}}} %
\renewcommand\omii[1]{{\color{black}{#1}}} %

\usepackage[sort,compress]{cite}
\usepackage{amsmath,amssymb,amsfonts}
\usepackage{algorithmic}
\usepackage{graphicx}
\usepackage{textcomp}
\usepackage{xcolor}
\usepackage{fancyhdr}
\usepackage{hyperref}

\def\BibTeX{{\rm B\kern-.05em{\sc i\kern-.025em b}\kern-.08em
    T\kern-.1667em\lower.7ex\hbox{E}\kern-.125emX}}

\title{\LARGE{\omf{\proposal: High-Performance, \omi{Energy-Efficient,} and Low-Cost\\ Metagenomic Analysis with In-Storage Processing}}}

\author{
Nika Mansouri Ghiasi$^1$ \hspace{0.5em} Mohammad Sadrosadati$^1$ \hspace{0.5em} Harun Mustafa$^1$ \hspace{0.5em} Arvid Gollwitzer$^1$ \vspace{0em} \\
Can Firtina$^1$ \hspace{0.5em} Julien Eudine$^1$ \hspace{0.5em} Haiyu Mao$^1$ \hspace{0.5em} Jo\"{e}l Lindegger$^1$ \hspace{0.5em} Meryem Banu Cavlak\omi{$^1$}\vspace{0em} \\
Mohammed Alser$^1$ \hspace{0.5em} Jisung Park$^2$ \hspace{0.5em} Onur Mutlu$^1$ \vspace{0em} \\
\normalsize{
$^1$ETH Zürich \hspace{0.5em} $^2$POSTECH
}
}

\begin{document}

\maketitle
\thispagestyle{firstpage}

\begin{abstract}
Metagenomics, \bback{the study of the genome sequences of diverse organisms in a common environment}, 
has led to significant advances in many 
fields.  
Since
the \hm{species} present in \hm{a} metagenomic sample are not known in advance, metagenomic analysis commonly involves the key tasks of determining the \hm{species} present in a sample and their relative abundances.  
These tasks require searching large metagenomic databases
\bback{containing information on different species’ genomes}. 
Metagenomic analysis suffers from significant data movement
overhead due to moving large amounts of low-reuse data from the storage system to the rest of the system.
In-storage processing 
can be a fundamental solution for reducing \omi{this} overhead. However, designing an in-storage processing system for metagenomics is challenging because existing approaches \omi{to \mganalysis} can\omi{not} be directly implemented in \new{storage}
effectively due to the hardware limitations of modern SSDs.

We propose \textbf{\proposal}, the \emph{first} in-storage processing system designed to significantly reduce the data movement overhead of \omf{the} end-to-end metagenomic analysis \omf{pipeline}. \proposal is enabled by our lightweight design that effectively leverages and orchestrates processing inside and outside the storage system.
Through our detailed analysis of the end-to-end metagenomic analysis pipeline and careful hardware/software co-design, we address in-storage processing challenges for metagenomics 
via specialized and efficient 
1)~task partitioning,
2)~data/computation flow coordination,
3)~storage technology-aware algorithmic optimizations,
\omi{4})~data mapping, and
\omi{5})~lightweight in-storage accelerators. 
\omii{\proposals design is flexible, capable of supporting different types of metagenomic input datasets, and can be integrated into various metagenomic analysis pipelines.}
Our evaluation shows that \proposal outperforms the state-of-the-art performance- and accuracy-optimized software metagenomic tools
by 2.7\omf{$\times$}--\new{37.2}$\times$ and 6.9\omf{$\times$}--\new{100.2}$\times$, respectively,  while matching the accuracy of the accuracy-optimized tool. \proposal achieves 1.5\omf{$\times$}--5.1$\times$ speedup compared to the state-of-the-art metagenomic hardware-accelerated  \omf{(using processing-in-memory)} tool, while achieving significantly higher accuracy.

\end{abstract}

\section{Introduction}
\label{sec:introduction}

Metagenomics\omi{,} an increasingly important domain in bioinformatics, \omi{requires the analysis of} the genome sequences of various organisms of \emph{different species} present in a common environment (e.g., human gut, soil, or oceans)~\cite{hhrlich2011metahit,sunagawa2015structure,fierer2017embracing}. 
\js{Unlike} traditional genomics~\cite{cali2020genasm,li2018minimap2,ham2020genesis,turakhia2018darwin,gupta2019rapid} \js{that} studies genome sequences \js{from} an individual \js{(}or a small group of individuals\js{)} of the \emph{same \omi{known} species}, metagenomics deals with \js{genome sequences whose species are \emph{not known} in advance in many cases, thereby requiring comparisons of} 
the \js{target} sequences against large databases of many reference genomes. 
Metagenomics has led to groundbreaking advances in many fields, such as precision medicine~\cite{kintz2017introducing,dixon2020metagenomics}, urgent clinical settings~\cite{taxt2020rapid}, understanding microbial diversity of an environment~\cite{afshinnekoo2015geospatial,hsu2016urban}, discovering early warnings of communicable diseases~\cite{john2021next,nagy2021targeted,nieuwenhuijse2017metagenomic}, and outbreak tracing~\cite{hadfield2018nextstrain}.
The pivotal role of metagenomics, together with rapid \js{improvements} in genome sequencing (e.g., reduced cost and improved throughput~\cite{berger2023navigating}), has resulted in \omiii{the} fast-growing adoption of metagenomics~\cite{dixon2020metagenomics,chiang2019from,chiu2019clinical}.

\omi{Given a metagenomic sample, a typical} workflow consists of three key steps\js{: \inum{i}~sequencing, \inum{ii}~basecalling, and \inum{iii}~\omi{\mganalysis}.}
First, \js{\omii{\emph{sequencing}} extracts} the genomic information of \emph{all} organisms in \omi{the} sample. 
\js{S}ince 
current sequencing \js{technologies \emph{cannot} process} \hhl{a} DNA molecule as a whole, \js{a sequencing machine} generate\js{s} randomly sampled, inexact fragments \js{of genomic information, called \emph{reads}}. \omi{A metagenomic sample contains organisms from several species, and during sequencing, it is unclear what species each read comes from.}
Second, \omii{\emph{basecalling}} converts the 
raw sequencer data of \js{reads} 
into 
sequence\hhl{s} of characters that represent \hhl{the} nucleotides\js{ \texttt{A}, \texttt{C}, \texttt{G}, and \texttt{T}}. 
\js{T}hird, \omi{\omii{\emph{\mganalysis}} identifies the distribution of different taxa (i.e., groups or categories in biological classification, such as species and genera) within a metagenomic sample. Metagenomic analysis commonly involves the two key tasks of} determin\omi{ing} the species \emph{present/absent} in the sample and \omi{finding} their \emph{relative abundances}.

To enable fast and efficient metagenomics for many critical applications, it is essential to improve the performance and energy efficiency of \omi{\mganalysis} due to \omii{at least} three \omii{major} reasons. 
First, \omi{\mganalysis} is typically performed much more frequently compared to the other two steps \js{ (i.e., sequencing and basecalling)} in the metagenomic workflow.
\js{While \omii{sequencing and basecalling}} are one-time tasks for a sample in many cases\js{, sequenced and basecalled reads in a sample} often \omii{need to} be analyzed \omii{over and over} in \emph{multiple studies} or \js{at} \emph{different times} in the same study~\cite{bokulich2020measuring}. 
Second, as shown in our motivational analysis in \sect{\ref{sec:motivation}} on a high-end server node, even \omi{when performing the \mganalysis step only once for a sample, this} step bottlenecks the end-to-end performance and energy \js{efficiency} of the workflow. 
Third, \js{the performance and energy-efficiency gaps between the \omi{\mganalysis} step and the other steps are expected to widen even more due to the rapid advances in sequencing \omii{and basecalling} technologies, such as}
significant increases in throughput and energy efficiency of sequencing~\cite{hu2021next,alser2020accelerating,katz2021sra,enastats,alser2022molecules,leinonen2010sequence} \omii{and basecalling\omiii{~\cite{singh2024rubicon,xu2021fast,lou2020helix,shahroodi2023swordfish,Samarakoon2023accelerated,cavlak2022targetcall}}}. \icut{(at rates higher than Moore's Law~\cite{berger2023navigating}), the development of sequencing technologies that enable analysis \omi{\emph{during}} sequencing~\cite{zhang2021real,firtina2023rawhash,kovaka2020targeted, mutlu2023accelerating,firtina2023rawhash2,lindegger2023rawalign}, and portable and energy-efficient high-throughput sequencers~\cite{minion21,jain2016oxford,cali2017nanopore}.} 
\js{For these reasons, \omi{simply} scaling up traditional systems \emph{cannot} effectively optimize the \mganalysis step enough to keep up with the rapid advances.}

\omi{Metagenomic} analysis suffers from significant data movement overhead \omii{because it requires} accessing large amounts of low-reuse data. Since we do not know the species present in a metagenomic sample, \omi{\mganalysis}  requires searching large databases (e.g., to several TBs\omii{~\cite{ncbi2023,karasikov2020metagraph,shiryev2023indexing,pebblescout,lemane2023kmindex,marchet2023scalable}} or more than a hundred TBs in emerging databases~\cite{shiryev2023indexing,pebblescout})
that contain information on different organisms' 
genomes. 
Database sizes are expected to increase further in the future, and at a fast pace.\footnote{For example, based on recently published trends, the ENA assembled/annotated sequence database \omi{size} currently \emph{doubles} every 19.9 months~\cite{enastats}, and the BLAST nt database \omi{size} doubled from 2021 to 2022~\cite{ntdouble}.} \omii{Two notable reasons for this growth are 1) the rapid evolution of viruses and bacteria~\cite{Lynch2010}, which necessitates frequent updates with new reference genomes~\cite{Nasko2018,o2016reference}, and 2) the fact that databases may include sequences from both highly curated reference genomes and from less curated metagenomic sample sets~\cite{karasikov2020metagraph,shiryev2023indexing}. Particularly, as over 99\% of Earth's microbes remain unidentified and excluded from curated reference genome databases~\cite{jiao2020microbial,Li2024}, the expanded databases improve sensitivity~\cite{Li2024}. Recent advances in the automated and scalable construction of genomic data from more organisms have further contributed to database growth by enabling the rapid addition of new sequences to databases~\cite{rautiainen2023telomere,jarvis2022semi}}.
Our motivational analysis (\sect{\ref{sec:motivation}}) of the state-of-the-art \mganalysis tools shows that data movement overhead from the storage system significantly impacts their end-to-end performance. Due to its low reuse, the data needs to move all the way from the storage system to the main memory and processing units for its first use, and \omi{it will} likely not be used  again \omii{or reused very little during} analysis. 
This \omi{unnecessary data movement}, \omiii{combined with} \omii{the} low computation intensity \omi{of \mganalysis} and \omii{the} limited I/O \omi{(input/output)} bandwidth, leads to large storage I/O overhead\omi{s for \mganalysis}.

While there ha\omi{s} been effort in accelerating \mganalysis, to our knowledge, no prior work fundamentally addresses its storage I/O overheads. Some works \omi{(e.g.,}\omii{~\cite{kim2016centrifuge,wood2019improved,muller2017metacache,song2024centrifuger,Dilthey2019,Fan2021}}\omi{)} aim to alleviate this overhead by applying sampling techniques to reduce the database size, but they incur accuracy loss, which is problematic for many use cases \omi{(e.g.,}\omiii{~\cite{milanese2019microbial,salzberg2016next,gihawi2023major,pockrandt2022metagenomic,berger2023navigating,Ackelsberg2015,Nasko2018,meyer2021critical,alser2020technology,alser2022molecules}}\omi{)}. 
Various other works (e.g.,~\cite{dashcam23micro, kobus2021metacache,wang2023gpmeta,wu2021sieve, hanhan2022edam,kobus2017accelerating,shahroodi2022krakenonmem,shahroodi2022demeter,armstrong2022swapping,jia2011metabing}) accelerate other bottlenecks in \mganalysis, such as computation and main memory bottlenecks. These works do not alleviate I/O overhead\omi{s}, whose impact on end-to-end performance becomes even larger \omii{(as shown in \sect{\ref{sec:motivation}})} when other bottlenecks \hhl{are} alleviated.

\js{\textbf{\omi{In-Storage Processing (ISP)}}, i.e., p}rocessing data directly in\js{side} the storage \js{device} where \js{target} data resides\js{,} can be a fundamental\omi{ly high-performance} approach to \js{mitigat\omii{ing}} the \js{data-movement bottleneck in \mganalysis, given its} three \omii{major} benefits. 
First, \js{ISP can significantly} reduce unnecessary data movement from\omii{/to the} storage system \js{by processing large amounts of low-reuse data inside the storage \omi{system} while sending only the results to the host}. 
Second, \js{ISP} can \js{leverage} \omii{each} \omi{SSD's\omi{\footnote{In this work, we focus on the predominant NAND flash-based SSD technology~\omii{\cite{cai2017error,cai-insidessd-2018}}. We expect that our insights and designs would benefit storage systems built with other emerging technologies.}} large internal} bandwidth \js{to access target data} without being restricted by the \omi{SSD's relatively smaller} external bandwidth. 
Third, \omi{ISP} \omii{alleviates} the overall \omi{execution} burden of applications \omi{with low data reuse} from the rest of the system (e.g., processing units and main memory), freeing up the host to perform other useful work \omi{instead}.

\omi{\textbf{Challenges of ISP.}} Despite \js{the} benefits \js{of ISP}, none of the existing approaches \js{\omiii{to} metagenomic \omiii{analysis}} can be  \omiii{effectively} implemented as an ISP system  
due to \js{the limited} hardware resources available in \omi{current} storage \js{devices}.
Some tools incur a large number of random accesses to search the database \omi{(e.g.,}\omii{~\cite{wu2021sieve,wood2019improved,wood2014kraken,truong2015metaphlan2,kim2016centrifuge,song2024centrifuger,ounit2015clark,piro2016dudes,Fan2021,piro2020ganon,Marcelino2020,milanese2019microbial}}\omi{)}, which hinders ISP's large potential by preventing the full utilization of  SSD internal bandwidth. This is due to costly conflicts in internal SSD resources (e.g., channels and NAND flash chips~\cite{nadig2023venice,tavakkol2018flin,kim2022networked}) caused by random accesses. 
Some tools predominantly incur more suitable streaming accesses \omi{(e.g.,}\omii{~\cite{lapierre2020metalign,pockrandt2022metagenomic,Dilthey2019,lemane2023kmindex,shen2022kmcp}}\omi{)}, 
but \js{doing so comes} at the cost of more computation and main memory capacity requirements that are 
challenging for ISP to meet due to \hhl{the} limited hardware resources available inside SSDs.
Therefore, directly adopting \omiii{either} approach \omi{(with random or streaming accesses)} in ISP incurs performance, energy, and storage device lifetime overheads.

\omi{\textbf{Our goal}} in this work is to improve metagenomic \omiii{analysis} performance by reducing \omi{the large} data movement overhead \omi{from the storage system} in a cost-effective manner. 
To this end, we propose \textbf{\proposal}, the \emph{first} ISP system designed to reduce the data movement overhead\omi{s inside} \omf{the} end-to-end \mganalysis pipeline.
The key idea of \proposal is to enable  \emph{cooperative} ISP for metagenomics, where we do not \omii{solely} focus on \omii{processing inside} the storage system \omii{but}, instead, \omii{we} capitalize on the strengths of processing \emph{both inside and outside} the storage system. \omii{We enable cooperative ISP} via a \omi{synergistic hardware/software co-design between the storage system and the host system}.

\textbf{\omi{Key Mechanism.}} We design \proposal as an efficient pipeline between the SSD and the host system to \inum{i}~\emph{leverage} and \inum{ii}~\emph{orchestrate} the capabilities of both. Based on our rigorous analysis of the end-to-end \mganalysis pipeline, we propose \omii{a} new hardware/software co-designed \omii{accelerator framework that consists of} five aspects.
First, 
we partition and map \omi{different parts of the \mganalysis pipeline} to the host and the ISP system such that each part \omii{is executed on the} most suitable architecture. 
Second, we coordinate the data/computation flow between the host and the SSD such that \proposal \inum{i}~completely overlaps the data transfer time between them with computation time to reduce the communication overhead between \omi{different parts}, \inum{ii}~leverage\hhl{s} SSD bandwidth efficiently, and \inum{iii}~does not require large DRAM inside the SSD or a large number of writes to the flash chips. 
Third, we devise storage technology-aware \omii{metagenomics} algorithm optimizations to enable efficient access patterns to the SSD. 
Fourth, we design lightweight in-storage accelerators to perform \proposal's ISP \omi{functionalities} while minimizing the required SRAM/DRAM buffer spaces inside the SSD. 
Fifth, we design an efficient data mapping scheme and Flash Translation Layer (FTL) specialized 
\hhl{to}
the characteristics of \omi{\mganalysis} to leverage the SSD's full internal bandwidth.

\omi{\textbf{Key Results.}} We evaluate \proposal with two different SSD configurations (performance-optimized~\cite{samsungPM1735} and cost-optimized~\cite{samsung870evo}).
We compare \proposal against three state-of-the-art \omi{software and hardware-accelerated} metagenomics tools: \inum{i}~Kraken2~\cite{wood2019improved}, which is optimized for performance, \inum{ii}~Metalign~\cite{lapierre2020metalign}, which is optimized for accuracy, and 
\inum{iii}~a state-of-the-art processing-in-memory accelerator, Sieve~\cite{wu2021sieve}, integrated into Kraken2 to accelerate its k-mer matching. 
\bback{By analyzing \emph{end-to-end} performance}, 
we show that
\proposal provides 2.7--37.2$\times$ and 1.5--5.1$\times$ speedup compared to Kraken2 and Sieve, respectively, while achieving significantly higher accuracy.
\proposal provides  6.9--100.2$\times$ speedup compared to Metalign, while providing the \omi{\emph{same}} accuracy \omii{(\proposal does not \omiii{affect analysis} accuracy compared to this accuracy-optimized baseline)}. 
\proposal provides large average energy reduction\hhl{s}
of 5.4$\times$ and 1.9$\times$ compared to Kraken2 and Sieve, respectively, and 15.2$\times$ compared to accuracy-optimized Metalign. \proposal's benefits come at a low area cost of 1.7\% \omi{over} the \omi{area of the} three cores~\cite{cortexr4} in a\omi{n} SSD controller~\cite{samsung860pro}. 

This work makes the following \omi{\textbf{key contributions}}:
\begin{itemize}[leftmargin=*, noitemsep, topsep=0pt]
    \item \omi{We demonstrate the end-to-end performance impact of storage I/O overheads in metagenomic \omiii{analysis}}.
    \item \omi{W}e propose \textbf{\proposal}, the \emph{first} \omii{in-storage processing (ISP)}  system tailored to reduce the data movement overhead of \omf{the} end-to-end \mganalysis pipeline,
    significantly reducing its I/O \omii{overheads} \omi{and improving its \omii{performance} and energy-efficiency}. 

    \item We present a new hardware/software co-design to enable an efficient and cooperative pipeline between the host and the SSD to \omii{alleviate I/O data movement overheads in} metagenomic \omii{analysis}.
    \item \omi{We rigorously evaluate \proposal and show that it} improves 
    performance and energy efficiency compared to the state-of-the-art metagenomics tools \omi{(software and hardware-accelerated)}, while maintaining high accuracy. It does so without relying on costly hardware resources throughout the system, making metagenomics more accessible for wider adoption. 
\end{itemize}

\section{Background}
\label{sec:background}

\subsection{Metagenomic Analysis}
\label{sec:background-metagenomics}

\begin{figure}[h]
         \centering
         \includegraphics[width=0.9\columnwidth]{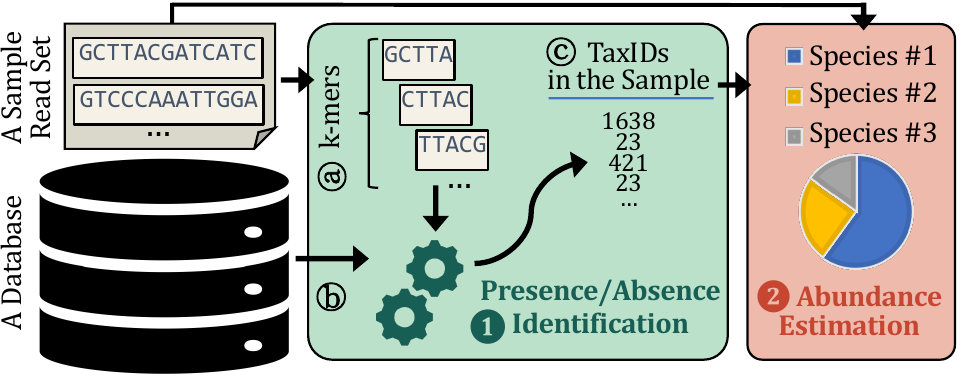}
         \caption{\irev{Overview of metagenomic analysis.}}
         \label{fig:mg-overview}
\end{figure}

\irev{\fig{\ref{fig:mg-overview}} shows\revid{\label{rev:A1}A1} an overview of metagenomic analysis,} which \omi{involves}
determin\omi{ing} the species \emph{present/absent} in the sample \irev{\circled{1}} and their \emph{relative abundances} \irev{\omi{(i.e.,} the relative frequencies \omii{of the occurrence of different species} in the sample\omi{)} \circled{2}}.

\subsubsection{Presence/absence Identification} 

To find species present in the sample, many tools \omi{(e.g.,\omii{~\cite{wu2021sieve,wood2019improved,wood2014kraken,truong2015metaphlan2,kim2016centrifuge,song2024centrifuger,ounit2015clark,piro2016dudes,Fan2021,piro2020ganon,Marcelino2020,milanese2019microbial,lapierre2020metalign,pockrandt2022metagenomic,Dilthey2019,lemane2023kmindex,shen2022kmcp}})} extract \emph{k-mers} (i.e., subsequences of length $k$) from the input quer\irev{ies in a sample} read \irev{set (\wcirc{a} in \fig{\ref{fig:mg-overview}})} and search for the k-mers in \irev{an input} \emph{reference database} \irev{(\wcirc{b})}. 
Each database contains k-mers extracted from reference genomes of a wide range of species. The database \rev{associates} each indexed k-mer with a \emph{taxonomic identifier (\revh{\omii{taxID})}}\footnote{A \omii{taxID} is an integer attributed to a cluster of related species.} of the \bback{reference} 
genome(s) the k-mer comes from. \omi{At the end of the presence/absence identification process, the metagenomic tool outputs the \omii{taxID}s of the species present in the sample} \irev{(\wcirc{c})}. 

The GB- or TB-scale databases typically support random (e.g.,\omii{~\cite{wu2021sieve,wood2019improved,wood2014kraken,truong2015metaphlan2,kim2016centrifuge,song2024centrifuger,ounit2015clark,piro2016dudes,Fan2021,piro2020ganon,Marcelino2020,milanese2019microbial}}) or streaming (e.g.,\omii{~\cite{lapierre2020metalign,pockrandt2022metagenomic,Dilthey2019,lemane2023kmindex,shen2022kmcp}}) access \bback{patterns}. 

\head{Tools with Random Access Queries (\randomio)}
\omi{Some tools (e.g.,\omii{~\cite{wu2021sieve,wood2019improved,wood2014kraken,truong2015metaphlan2,kim2016centrifuge,song2024centrifuger,ounit2015clark,piro2016dudes,Fan2021,piro2020ganon,Marcelino2020,milanese2019microbial}})} commonly perform random accesses to search their database.
A state-of-the-art tool in this category is Kraken2~\cite{wood2019improved}, which maintains a hash table \omii{that maps} each indexed k-mer to a \omii{taxID}.  To identify which species are present in a set of queries, Kraken2 extracts k-mers from the read queries and searches the \omii{hash table} to retrieve the k-mers' associated \omii{taxID}s. 
For each read, Kraken2 collects the \omii{taxID}s of that read's k-mers and, based on the occurrence frequencies of these \omii{taxID}s, uses a classification algorithm to assign a single \omii{taxID} to each read. Finally, Kraken2 identifies the species \omii{present in the sample based on the \omii{taxID}s of the reads in the sample.}

\head{Tools with Streaming Access Queries (\streamio)}
Some tools (e.g.,\omii{~\cite{lapierre2020metalign,pockrandt2022metagenomic,Dilthey2019,lemane2023kmindex,shen2022kmcp}}) predominantly feature streaming accesses to their databases. A state-of-the-art tool in this category is Metalign~\cite{lapierre2020metalign}. Presence/absence identification in Metalign is done via 1) preparing the input read set queries, and 2) finding species present in them. To process the queries, the tool extracts k-mers from the reads and sorts them. Finding the species in the sample involves two steps.  First, \omii{the tool} finds the \omii{\emph{intersecting k-mers}, which are k-mers that are common} between the query k-mers and a pre-sorted reference database. In this step, the tool uses large k-mers (e.g., $k=60$) for both the queries and the database to maintain a low false positive rate. This is because large k-mers are more unique, and matching a long k-mer ensures that the queries have at least one long and specific match to the database. Second, the tool finds the \omii{taxID}s of the intersecting k-mers by searching for the intersecting k-mers or \hm{their} prefixes in a smaller \emph{sketch database} of variable-sized k-mers. Each \emph{sketch} is a small representative subset of k-mers associated with a given \omii{taxID}. Searching for \omii{both the intersecting k-mers and their} prefixes in this step increases the true positive rate \omii{(i.e., \omiii{species} correctly identified \omiii{as} present in the sample out of all species \omiii{actually} present in the sample)} by expanding the number of matches.

\subsubsection{Abundance Estimation}

After finding the \omi{\omii{taxID}s of the} species \omi{present} in the sample, some applications require a more sensitive step to find the species' relative abundances\omii{~\cite{sun2021challenges,lu2017bracken,koslicki2016metapalette,piro2016dudes,truong2015metaphlan2,shen2022kmcp,Fan2021,lapierre2020metalign,kim2016centrifuge,song2024centrifuger,dimopoulos2022haystac}} \irev{in the sample}. Different tools implement their own approaches for estimating abundances, from lightweight statistical models\omii{~\cite{lu2017bracken,dimopoulos2022haystac,koslicki2016metapalette}} to more accurate but computationally-intensive read mapping\omii{~\cite{lapierre2020metalign,kim2016centrifuge,milanese2019microbial,piro2016dudes,Fan2021,truong2015metaphlan2}}. 
\bback{Read mapping is the process of finding potential matching locations of reads against \omii{one or more} reference genomes}. 
Metagenomic tools can map the reads against reference genomes of species in the sample,
accurately determining the number of reads belonging to each species.

\subsection{SSD Organization}
\label{sec:background-ssd}

\begin{figure}[b]
         \centering
         \includegraphics[width=0.8\columnwidth]{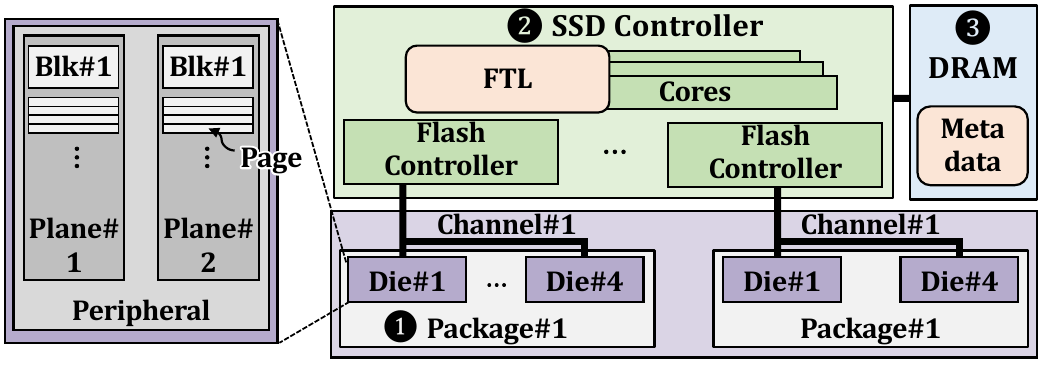}
         \caption{Organizational overview of a modern SSD.}
         \label{fig:ssd}
\end{figure}

\fig{\ref{fig:ssd}} depicts the organization of a modern NAND flash-based SSD, \omii{which} consist\omii{s} of three main components. 

\head{\circled{1}~NAND Flash Memory}
A NAND package consists of multiple \emph{dies} or \emph{chips}, sharing the NAND package's I/O pins. 
One or multiple packages share a command/data bus or \emph{channel} to communicate with the SSD controller.
Dies can operate independently, but each channel can be used by only one die at a time to communicate with the controller.
\bback{Each die has multiple (e.g., 2 or 4) \emph{planes}, each with thousands of \emph{blocks}.
Each block has hundreds to thousands of 4--16 KiB \emph{pages}.
NAND flash memory performs read/write operations at page granularity but erase operations at block granularity.
The peripheral circuitry to access pages is shared among the planes in each die. Hence, it is possible for the planes in a die to operate concurrently when accessing pages (or blocks) at the same offset. This mode is called the \emph{multiplane} operation}.

\head{\circled{2}~SSD Controller}
An SSD controller consists of two key components. First, multiple cores run the FTL,
which is responsible for communication with the host, internal I/O scheduling, and various SSD management tasks. Second, per-channel hardware flash controllers manage request handling\omii{~\cite{tavakkol2018flin,cai2013program}} and error correction for the NAND flash chips\omiii{~\cite{dong2010use,zhao2013ldpc,bose1960class,wang2014enhanced,cai-hpca-2017, luo2018improving,luo-hpca-2018,cai2015read,cai2013error, cai2012flash, cai2017error, ha2015integrated, cai-insidessd-2018,luo2015warm}}.

\head{\circled{3}~DRAM} Modern SSDs use low-power DRAM\omii{~\cite{zou2022assasin}} to store metadata for SSD management tasks. Most of the \hm{DRAM} capacity inside the SSD (i.e., internal DRAM) is used to store the logical-to-physical (\omii{i.e.,} L2P) mappings\bback{, which are typically maintained at a granularity of 4KiB to enhance random access performance. \omii{I}n a 32-bit architecture, \omii{with} 4 bytes of metadata stored for every 4KiB of data,} 
\omii{the required capacity for} L2P \omii{mappings} is about 0.1\% of the SSD's capacity. For example, a 4-GB LPDDR4 DRAM is used for a 4-TB SSD~\cite{samsung860pro}.

\subsection{\irev{In-Storage Processing}\revid{\label{rev:D1.1}D1.1}}
\label{sec:background-isp}

\irev{Processing data directly in storage via \emph{in-storage processing} (ISP) can be a fundamental\omi{ly high-performance} approach for reducing the overheads of moving large \omii{amounts of} low-reuse data across the system, providing three key benefits. First, \omi{ISP} reduces unnecessary data movement from \omi{the} storage \omi{system}. Second, \omi{ISP} reduces the \omi{execution} burden \omi{of applications with low data reuse from the rest of} the system, allowing it to perform other useful tasks. Third, as shown by many prior works \omiii{(e.g.,~\cite{li2023ecssd,mailthody2019deepstore,kang2021mithrilog,koo2017summarizer,mansouri2022genstore,wang2024beacongnn,jang2024smart,kim2023optimstore,li2021glist})}, ISP can benefit from the SSD's internal bandwidth. 
In\revid{\label{rev:A4.1}A4.1} modern SSDs, the internal bandwidth is usually larger than the external. For example, a modern SSD controller~\cite{anandcontroller} supports \omii{6.5 G}B/s external
bandwidth and 19.2 GB/s internal bandwidth (16 channels \omii{with a} maximum per-channel bandwidth of 1.2 GB/s). It is essential to overprovision the internal bandwidth to avoid hurting the user-perceived external I/O bandwidth by reducing the negative impact of 1) channel conflicts~\cite{nadig2023venice,kim2023decoupled,kim2022networked,tavakkol2014design} and 2) the SSD’s internal data migration for management tasks such as garbage collection~\cite{tavakkol2018flin,kim2020evanesco,cai2017error,park-dac-2016, park-dac-2019}, wear-leveling~\cite{chang2007efficient,cai2017error,cai-insidessd-2018}, and data refresh~\cite{cai2013error,luo2018improving, cai2017error,cai2012flash}}.

\irev{Some prior works propose ISP systems in the form of \emph{special-purpose} accelerators for different applications\omii{~\cite{liang2019cognitive,kim2020reducing,lim2021lsm,li2021glist,wang2016ssd,lee2020neuromorphic,kang2021s,han2021flash,wang2022memcore,wang2018three,han2019novel,choi2020flash,mailthody2019deepstore,pei2019registor,jun2018grafboost, do2013query, seshadri2014willow,kim2016storage, riedel2001active,riedel1998active,lee2022smartsage,jeong2019react, jun2016storage,li2023ecssd,wang2024beacongnn,jang2024smart}}. 
Several prior works propose \emph{general-purpose} processing inside storage devices\omii{~\cite{gu2016biscuit, kang2013enabling, wang2019project,acharya1998active,keeton1998case,riedel1998active,riedel2001active,merrikh2017high,tiwari2013active,tiwari2012reducing,boboila2012active,bae2013intelligent,torabzadehkashi2018compstor,kang2021iceclave,zou2022assasin}}, bulk-bitwise operations using \omii{NAND} flash~\cite{gao2021parabit,park2022flash}, or SSDs in close integration with FPGAs~\cite{jun2015bluedbm, jun2016bluedbm, torabzadehkashi2019catalina, lee2020smartssd, ajdari2019cidr, koo2017summarizer} or GPUs~\cite{cho2013xsd}}.

\section{Motivational Analysis}
\label{sec:motivation}

\subsection{Criticality of Metagenomic Analysis}
\label{sec:motivation-criticality}

\irev{\omi{By} enabl\omi{ing the analysis of} the genomes of organisms \omi{from} different species in a common environment, metagenomics overcomes \omii{a} limitation of traditional genomics, \omi{which} requires culturing individual known species in isolation. \omi{This limitation} has been a major roadblock in many clinical and environmental use cases\hhl{~\cite{national2007new}}. The impact of metagenomics has been rapidly increasing in many areas that each have broad implications for society, such as health\hhl{~\cite{kintz2017introducing,taxt2020rapid}}, agriculture\hhl{~\cite{cdcpulsenet}}, environmental monitoring\hhl{~\cite{john2021next,nagy2021targeted,nieuwenhuijse2017metagenomic,hadfield2018nextstrain}}, and many other critical areas.  Due to its importance, metagenomics has attracted wide global attention, with medical and government health institutions heavily investing in metagenomic analysis\hhl{~\cite{downie2023surveillance,cdcpulsenet,cdcamd}}. The global amount of genomic data that is incorporated in metagenomic workflows is growing \omii{exponentially}\hhl{\omii{~\cite{stephens2015big,Nasko2018}}}, doubling every several months\hhl{\omii{~\cite{CheckHayden2015,enastats,ntdouble,Nasko2018}}}, and is projected to surpass \omii{the data growth rate of} YouTube and Twitter\hhl{~\cite{wu2021sieve,stephens2015big,CheckHayden2015}}}.

\irev{In metagenomics,} the analysis step bottlenecks the end-to-end performance of the workflow, \irev{and therefore,} poses a pressing need for acceleration~\cite{wu2021sieve,hanhan2022edam,chiang2019from,edgar2022petabase,taxt2020rapid,sereika2022oxford} for three reasons. First, the sequencing and basecalling steps for a sample are \omii{usually} one-time tasks\omii{~\cite{wang2021nanopore,hu2021next,alser2022molecules}}. \omii{I}n many cases, the reads from a single sequenced sample can be analyzed by \emph{multiple studies} or \emph{at different times} in the same study. \irev{This is because\revid{\label{rev:E1.4}E1.4}} 
\irev{\inum{i}~there are many heuristics involved in metagenomics, and achieving a desired sensitivity-specificity tradeoff \icut{(e.g., in accuracy-critical medical or environmental use cases)} commonly requires parameter tuning\hhl{~\cite{bokulich2020measuring}}, or using different databases \omii{created} with different parameters or genomes\hhl{~\cite{schuele2021future}}, and
\inum{ii} a sample can be analyzed several times with databases that are regularly updated with new genomes, or with syndrome-specific targeted databases\hhl{~\cite{schuele2021future}}}.
Second, even \omi{when performing the metagenomic analysis step only once for a sample,} the \emph{throughput} of \omi{this step} is significantly lower than the sequencing \revid{\label{rev:E1.1}E1.1}throughput of modern sequencers (e.g.,~\cite{illuminax}). \irev{While sequencing one sample can take a long time, } a single sequencing machine can sequence \emph{many samples} from different sources \icut{(e.g., different patients or environments)}in parallel~\cite{hu2021next,shokralla2015massively}, achieving very high throughput.  Our analysis with a state-of-the-art metagenomic tool~\cite{lapierre2020metalign} shows that analyzing the data, sequenced and basecalled by a sequencer in 48 hours, takes 38 days on a high-end server node (detail\omii{ed configurations} in \sect{\ref{sec:methodology}}). Such long analysis poses serious challenges, specifically for time-critical use cases (e.g., clinical settings~\cite{taxt2020rapid} and timely \icut{and widespread}surveillance of infectious diseases~\cite{hadfield2018nextstrain}). Since the growth rate of sequencing throughput is higher than Moore's Law~\cite{berger2023navigating},
this already large gap between sequencing and \omi{analysis} throughput is widening~\cite{hu2021next,katz2021sra,enastats,leinonen2010sequence}, and \omi{simply} scaling up traditional systems for analysis is not efficient. 
\omii{Third, the development of sequencing technologies that enable analysis \omi{\emph{during}} sequencing~\cite{zhang2021real,firtina2023rawhash,kovaka2020targeted, mutlu2023accelerating,Payne2021,Bao2021,ulrich2022readbouncer} increasingly \omiii{necessitates the need for} fast analysis \omiii{that can keep up with sequencing throughput}}.

\omii{T}he analysis step is \omii{also} the primary energy bottleneck \bback{in the metagenomic workflow}, 
and optimizing its efficiency is vital as sequencing technologies rapidly evolve. For example, a high-end sequencer~\cite{illuminax} uses 405 KJ to sequence and basecall 100 million reads, with 92.5 Mbp/s throughput and 2,500 W power consumption~\cite{illuminax}. In contrast, processing this dataset on a commodity server (\omii{detailed configurations in} \sect{\ref{sec:methodology}}) requires 675 KJ, accounting for 63\% of total energy. The need to enhance the analysis' energy efficiency is further increasing for two reasons.
First, sequencing efficiency has \omi{been} \omiii{continually} improv\omi{ing}. For example, a new version of Illumina sequencer from 2023~\cite{illuminax} provides 44$\times$ higher throughput at only 1.5$\times$ \omii{higher} power consumption compared to an older version~\cite{illumina} from 2020, resulting in much better sequencing energy efficiency. Therefore, \omi{simply} relying on scaling up commodity systems to improve the analysis throughput worsens the energy bottleneck. Second, the increased adoption of compact \emph{portable sequencers}~\cite{jain2016oxford} for on-site metagenomics (e.g., in remote locations~\cite{pomerantz2018real} or for personalized bedside care~\cite{chiang2019from}) offers high-throughput sequencing with low energy costs. This further amplifies the need for energy- and cost-effective analysis that can match the portability and convenience of these 
sequencers.

\icut{Due to{\setstretch{0.5}\omn{\tiny Removing this revision response. The points are covered in the next section. We just put this in revision so that it summarizes some points at the same place for the reviewer.}} its importance\revid{\label{rev:B1.2}B1.2\\------\\\label{rev:E1.5}E1.5}, there have been great efforts in accelerating metagenomic analysis. However, to our knowledge, no prior work fundamentally addresses its  I/O overheads. Some works~\cite{kim2016centrifuge,wood2019improved,muller2017metacache} apply sampling to shrink databases, sacrificing accuracy, which affects many applications~\cite{milanese2019microbial,salzberg2016next,gihawi2023major,pockrandt2022metagenomic}. Several works~\cite{dashcam23micro, wu2021sieve,kobus2021metacache,wang2023gpmeta, hanhan2022edam,kobus2017accelerating,shahroodi2022krakenonmem,shahroodi2022demeter,armstrong2022swapping,jia2011metabing} target computation and \omi{main} memory \omi{bottlenecks} in metagenomic analysis but neglect I/O overhead, which becomes more critical as other bottlenecks are mitigated.}

\subsection{Data Movement Overheads}
\label{sec:motivation-ovhd}

\bback{We conduct experimental analysis to assess the storage system's impact on the performance of metagenomic analysis}.

 \head{Tools and Datasets} We analyze two state-of-the-art  tools 
 for presence/absence identification: 
1)~Kraken2~\cite{wood2019improved}, which queries \hhl{its} large database with random access patterns (\randomio),%
\footnote{We experiment with both techniques of accessing the database devised in the \randomio baseline~\cite{wood2014kraken} and report the best timing. The first technique uses mmap to access the database, while the second technique loads the entire database from the SSD to DRAM as the first step when the analysis starts. In this experiment, the second approach performs slightly better since, when analyzing our read set, the application accesses most parts of the database.}
and 2)~Metalign~\cite{lapierre2020metalign},
which exhibits \omii{mostly} sequential streaming accesses to \hhl{its} database (\streamio). We use the best-performing thread count for each \omi{tool}.
We use a query sample with 100 million reads
(CAMI-L, detailed in \sect{\ref{sec:methodology}}) from the CAMI dataset~\cite{meyer2021critical}, commonly used for profiling metagenomic tools. 
We generate a database based on \hm{microbial genomes drawn from NCBI's databases}~\cite{ncbi2020,lapierre2020metalign} using default parameters for each tool. For Kraken2~\cite{wood2019improved}, this results in a 293 GB database. For Metalign~\cite{lapierre2020metalign}, this results in a 701 GB k-mer database
and \gram{a} 6.9~GB sketch tree. To show the impact of \omi{database size}, we also analyze larger k-mer databases (0.6~TB and 1.4~TB for Kraken and Metalign, respectively) that include more species.

\head{System Configurations} 
We use a high-end server with \hhl{an} AMD EPYC 7742 CPU~\cite{amdepyc} and 1.5-TB DDR4 DRAM\omii{~\cite{ddr4sheet}}. \omii{We n}ote that the DRAM size is larger than the size of all data accessed during the analysis by each tool. This way, we can analyze the fundamental I/O overhead of moving large amounts of low-reuse data from storage to the main memory without being limited by DRAM capacity.
We evaluate I/O \omii{overheads using}:
1)~a cost-optimized SSD (\ssdc)~\cite{samsung870evo} with a SATA3 interface~\cite{SATA}, 
2)~a performance-optimized SSD (\ssdp)~\cite{samsungPM1735} with a PCIe Gen4 interface~\cite{PCIE4}, and
3)~a hypothetical configuration with zero performance overhead due to storage I/O (\dram). 
\bback{\ssdp provides an order-of-magnitude higher sequential-read bandwidth than \ssdc \omii{~(detailed configurations in Table~\ref{table:SSD_config})}. 
However, scaling up storage capacity only using performance-optimized SSDs is challenging due to their much higher prices \omiii{(e.g.,~\cite{PM1735price,PM9A3price,EVO870price})} and fewer PCIe slots compared to SATA \omii{slots available} on servers \omiii{(e.g.,~\cite{amdepyc})}}.

\begin{figure}[t]
    \centering
    \includegraphics[width=0.9\linewidth]{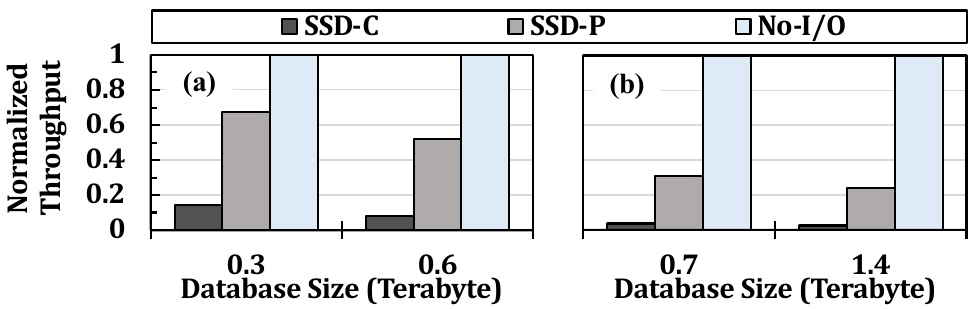}
    \caption{Performance of (a) \randomio and (b) \streamio under different storage configurations and database sizes.}
    \label{fig:motivation}
\end{figure}

\head{Results and Analysis}
\fig{\ref{fig:motivation}} shows the performance \omii{(throughput in terms of \#queries/sec)} of the tools normalized \omiii{to} \dram. We make three key observations. First, I/O overhead has a large impact on
performance \bback{for all cases}.
Compared to \ssdc (\ssdp), \dram leads to 9.4$\times$ (1.7$\times$) and 32.9$\times$ (3.6$\times$) better performance in \randomio and \streamio (average\omi{d} across both databases), respectively.   
While both baselines significantly suffer from large I/O overhead, we observe a relatively larger impact on \streamio due to its lower data reuse compared to \randomio. \omii{This is because the lower the data reuse, the less effectively the initial I/O cost can be amortized.}
Second, even using the \omii{costly} state-of-the-art SSD (\ssdp) does not alleviate this overhead, leaving large performance gaps between \ssdp and \dram in both tools.    
Third, I/O overhead increases as the databases grow. 
For example, in \randomio, the performance gap between \ssdc and \dram widens from 7.1$\times$ to 12.5$\times$ as the database expands from 0.3 TB to 0.6 TB. 
\bback{Based on these observations, We conclude that I/O accesses lead to large overheads in metagenomic \omii{analysis}, an issue expected to \omii{worsen} in the future}.

This I/O overhead, stemming from the need to move large amounts of low-reuse data\hhl{,} is a fundamental problem \omii{that is} hard to avoid. One might think it is possible to avoid this overhead by 1) \omii{using} sampling\revid{\label{rev:B6}B6} \irev{techniques} to shrink database \omii{sizes} \omi{(e.g.,}\omii{~\cite{kim2016centrifuge,wood2019improved,muller2017metacache,song2024centrifuger,Dilthey2019,Fan2021}}\omi{)} or 2) keeping all data required by metagenomic analysis \omii{completely and always} resident in main memory. Neither of these solutions is suitable. The first approach inevitably reduces accuracy\omii{~\cite{berger2023navigating,milanese2019microbial,meyer2021critical}} to levels unacceptable for many 
use cases \omi{(e.g.,}\omii{~\cite{milanese2019microbial,salzberg2016next,gihawi2023major,pockrandt2022metagenomic,berger2023navigating,Ackelsberg2015,Nasko2018,meyer2021critical}}\omi{)}.
The second approach is energy inefficient, costly, unscalable, and unsustainable due to two reasons. First, the sizes\revid{\label{rev:E1.3}E1.3} of metagenomic databases (which are already large, i.e., in some recent examples, exceeding a hundred terabytes~\cite{shiryev2023indexing,pebblescout}) have been increasing rapidly.
For example, recent trends show the \emph{doubling} of different important databases in only several months\omii{~\cite{enastats,ntdouble,Nasko2018}}. Second, 
regardless of \hhl{the} sizes\revid{\label{rev:E1.2}E1.2} \irev{of \omii{individual} databases}, different analyses need \emph{different databases}, \irev{with information from different sets of genomes or with varying parameters}. For example, a medical center may use various databases for its patients \irev{based on the patients' conditions\hhl{~\cite{schuele2021future}}} (e.g., for different viral infections~\cite{centrifuge_db,kim2016centrifuge}, sepsis~\cite{taxt2020rapid}, etc.).
Therefore, it is inefficient and unsustainable to maintain \emph{all} data required by \emph{all possible} analyses in DRAM at all times.\footnote{Ultimately, these are the same reasons that the metagenomic\omii{s} community has been investigating storage efficiency (e.g., the \omii{afore}mentioned sampling techniques\omii{~\cite{kim2016centrifuge,wood2019improved,muller2017metacache,song2024centrifuger,Dilthey2019,Fan2021}}) as opposed to merely relying on scaling the system's \omiii{main memory}~\cite{karasikov2020metagraph,pockrandt2022metagenomic,lemane2022kmtricks,alanko2023themisto,fan2023fulgor}.}

The I/O impact on end-to-end performance becomes even more prominent in emerging systems in which other bottlenecks \icut{(e.g., in computation or main memory)} are alleviated. 
For example, while metagenomics can benefit from near-data processing at the main memory level, i.e., processing-in-memory (PIM)\omiii{~\cite{wu2021sieve,shahroodi2022krakenonmem,shahroodi2022demeter,hanhan2022edam,zou2022biohd,mutlu2022modern,ghose2019processing,mutlu2019processing,ghose2018enabling}}, these approaches still incur the overhead of moving the large, low reuse data from \omi{the} storage \omi{system}. In fact, by \omii{alleviating} other \omii{bottlenecks}, the impact of I/O on end-to-end performance increases. For example, for the 0.3-TB and 0.6-TB Kraken2 databases, 
\hhl{using}
a state-of-the-art PIM accelerator~\cite{wu2021sieve} of Kraken2,  \dram is on average 26.1$\times$ (3.0$\times$) faster than \ssdc (\ssdp). We conclude that while \omii{accelerating other bottlenecks in
metagenomic analysis (e.g., main memory bottlenecks)} can provide significant benefits, \omiii{doing so} does not alleviate the overhead\omii{s} of moving large, low-reuse data from the storage \omiii{system}.

\subsection{Our Goal}
\label{sec:motivation-goal}

\irev{ISP 
can be a fundamental solution for reducing data movement. However,} designing an ISP system for metagenomics is challenging because none of the existing approaches can be directly implemented as an ISP system effectively due to \hhl{an} SSD's constrained hardware resources. 
\bback{Techniques such as \randomio hinder leveraging ISP’s large potential by preventing the full utilization of the SSD’s internal bandwidth due to costly conflicts in internal SSD resources~\cite{nadig2023venice,tavakkol2018flin,kim2022networked} \omii{caused by random accesses}. Techniques such as \streamio predominantly incur more suitable streaming accesses, but at the cost of more computation and main memory capacity requirements, posing challenges for ISP.
Therefore, directly adopting \omii{existing metagenomic analysis} approaches in storage incurs performance, energy, and lifetime overheads}.
\omi{\textbf{Our goal} in this work is to improve the performance \omiii{and efficiency} \omii{of metagenomic analysis} by reducing \omi{the large} data movement overhead \omi{from the storage system} in a cost-effective manner}.

\section{\proposal}
\label{sec:mechanism}

\omii{W}e propose \proposal, the \emph{first} ISP system \omi{designed \omiii{for the end-to-end \mganalysis pipeline} to reduce \omiii{its} data movement overhead\omi{s \omii{from the storage system}}}.
\irev{\proposal is primarily designed as a system for accelerating\revid{\label{rev:D3.1}D3.1} metagenomic analysis}. \proposal extends the existing SSD controller and FTL \irev{\omii{\emph{without}} impacting the baseline SSD functionality. Therefore, when metagenomic acceleration is not in progress,} the SSD can be accessible for all other applications\irev{, similar to a general-purpose SSD}. 

We address the challenges \omi{of ISP for \mganalysis} via hardware/software co-design to enable \omii{what we call} \revid{\label{rev:D1.2}D1.2}\emph{cooperative ISP}. \omii{In other words, we do not solely focus on processing inside the storage system but, instead, we \omiii{exploit} the strengths of processing both inside and outside the storage system}. \irev{\proposal enables} an efficient pipeline between the host \omi{system} and \js{the storage system} to \js{maximally leverage and orchestrate the capabilities \omii{of both systems}}.

\irev{It\revid{\label{rev:D1.3}D1.3} is possible for MegIS’s ISP \omi{steps} to run on our lightweight specialized ISP accelerators or, alternatively, on the existing embedded cores in the SSD controller\footnote{\irev{These cores are available for \proposals ISP\revid{\label{rev:D3.3}D3.3}  since we envision that during metagenomic acceleration, \proposal is not used as a general-purpose SSD and does not run the baseline FTL. Instead, it runs \omii{\proposal{}~FTL}, which only performs lightweight and infrequent tasks during ISP \omii{(see \sect{\ref{sec:mech-ftl}})}.}} or other general-purpose ISP systems (e.g., \cite{torabzadehkashi2019catalina,zou2022assasin,jun2015bluedbm,gu2016biscuit}). 
This is because, leveraging our optimizations, \proposals ISP \omi{steps} require only simple computation and small buffers.
Efficiently performing metagenomics on any of these underlying hardware units requires \proposals specialized task partitioning, data/computation flow coordination, storage technology-aware algorithmic optimizations, and data mapping. The ability to leverage existing hardware units (embedded SSD cores or general-purpose ISP \omi{systems}) \omii{helps with} \proposals\revid{\label{rev:E1.6}E1.6\\------\\\label{rev:B1.4}B1.4} ease of adoption. Ultimately, choosing between \omii{different} MegIS configurations (our specialized lightweight \omi{accelerators} or general-purpose hardware) is a design decision \omiii{that has} \omii{various} tradeoffs, with specialized accelerators achieving \omii{the highest} performance and power efficiency (\sect{\ref{sec:eval-main}})}. 

\subsection{\irev{Overview}\revid{\label{rev:A2}A2}}
\label{sec:mech-overview}

\begin{figure}[t]
    \centering
    \includegraphics[width=0.85\linewidth]{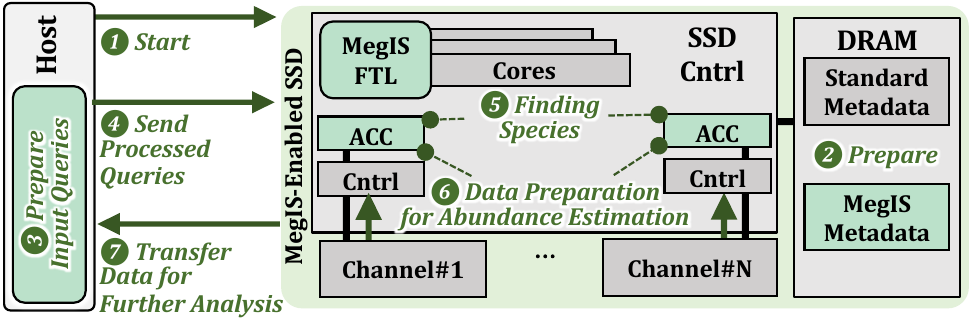}
    \caption{Overview of \proposal.}
    \label{fig:metastore_overview}
\end{figure}

\fig{\ref{fig:metastore_overview}} shows \irev{an overview of} \proposal's \irev{steps.
We design \proposal as an efficient pipeline in the SSD and the host system.
We develop \omii{\proposal{}~FTL} \omi{(\sect{\ref{sec:mech-ftl}})}, which is responsible for communication with the host system and data flow across the SSD hardware components (e.g., NAND flash chips, internal DRAM, and hardware accelerators) when running metagenomic analysis.}
Upon receiving \gram{a} notification from the host to initiate metagenomic analysis (\circled{1} \irev{in \fig{\ref{fig:metastore_overview}}}), \proposal readies itself 
by loading the necessary \omii{\proposal{}~FTL} metadata
(\circled{2}).
\irev{After this preparation, \proposal starts its three-step execution.} 
In \rev{\textbf{Step 1} \omi{(\sect{\ref{sec:mech-stage1}})}, the host processes the input read queries (\circled{3})} and transfers them in batches to the SSD (\circled{4}). 
In \textbf{Step 2} \omi{(\sect{\ref{sec:mech-stage2}})}, the ISP units \omii{(ACC in \fig{\ref{fig:metastore_overview}})} find the species present in the sample (\circled{5}). Steps 1 and 2 run in a pipelined manner.
In \textbf{Step 3} \omi{(\sect{\ref{sec:mech-stage3}})}, \omii{\proposal{}} prepar\omii{es} (\circled{6}) and transfer\omii{s} (\circled{7}) the data needed for any further analysis. \omii{By doing so, \proposal facilitates integration with different abundance estimation approaches.}
\irev{\proposal leverages the SSD's \emph{full internal\revid{\label{rev:A4.2}A4.2} bandwidth} since it avoids channel conflicts (due to its specialized data/control flow) and frequent management tasks (by not requiring writes during its ISP steps)}.

\subsection{Step 1: Preparing the Input Queries}
\label{sec:mech-stage1}

\omi{In this step, \proposal prepares the input read queries in a metagenomic sample for metagenomic analysis}. \proposal works with lexicographically-sorted data structures to avoid expensive random accesses to the SSD (similar to \streamio{}, described in \sect{\ref{sec:background-metagenomics}}). Like many other metagenomic tools (\omi{e.g.,}\omii{~\cite{wood2014kraken,shahroodi2022krakenonmem,lapierre2020metalign,truong2015metaphlan2,kim2016centrifuge,song2024centrifuger,ounit2015clark,Dilthey2019,koslicki2016metapalette,shen2022kmcp,Marcelino2020,piro2016dudes,Fan2021,piro2020ganon,wood2019improved}}), we assume the sorted k-mer \emph{databases} are pre-built before the analysis. However, sorting k-mers extracted from the \emph{input query read set} 
is inefficient \omii{to perform \omii{offline}} due to the need to store a large data structure (sorted k-mer set) with each sample, potentially larger than the sample itself, causing significant storage capacity waste.
Therefore, to prepare the input queries, \proposal 1) extracts k-mers from the sample \omi{(\sect{\ref{sec:mech-stage1-kmer-extraction}})}, 2) sorts the k-mers \omi{(\sect{\ref{sec:mech-stage1-sorting}})}, and if needed, 3) prunes some k-mers \hhl{according to} user-defined criteria \omi{(\sect{\ref{sec:kmer-exclusion}})}.

We execute this step in the host \omi{system} for three reasons. First, \omi{this step} benefits from the relatively larger DRAM and \omii{more powerful} comput\omii{ation} resources in the host. Second, due to the 
large \omiii{host-side} DRAM, performing this step in \gram{the} host leads to significantly fewer writes to the flash chips\omii{, positively impacting}
lifetime. For typical metagenomic read sets, storing k-mers extracted from reads within a sample takes tens of gigabytes (e.g., on average 60 GB with standard CAMI read sets \cite{meyer2021critical}). 
While generating and sorting k-mers inside the SSD 
is possible, it would necessitate frequent writes to flash chips \omii{or much larger DRAM}. Third, by leveraging the host \omi{system} for this step, we enable pipelining and overlapping Step 1 with Step 2 (which searches the large, low-reuse database).

\bback{To efficiently execute \omii{Step 1} on the host \omi{system}, we need to ensure two points. First, partitioning the application between the host system and the SSD should not incur significant overhead\omi{s} due to data transfer time. 
Second, while it is reasonable in most cases to expect the host DRAM to be large enough to contain all extracted k-mers from a sample, \proposal should accommodate scenarios where this is not the case and minimize the performance, lifetime, and endurance overheads of writes \omii{to flash chips} due to page swaps \omii{(i.e., moving data back-and-forth between the host DRAM and the SSD when the host DRAM is smaller than the application's working set size)}}.

\irev{\omii{The sequences in \proposals databases are encoded with two bits per character (i.e., \texttt{A}, \texttt{C}, \texttt{G}, \texttt{T} in DNA alphabet)} during their offline generation. For the read sets, \proposal is able to work with different formats. We \omii{perform} the first analysis step \omii{(Step~1)} in the host system so that any format conversion can be flexibly incorporated there (e.g., from ASCII \omii{or binary} to 2-bit encoding). 
\omii{The overhead of format conversion is negligible since it involves a straightforward transformation of the four nucleotide bases to the 2-bit encoded format}.
For the remainder of \proposals pipeline, we use \omiii{the} 2-bit encoding. 
}

\subsubsection{K-mer Extraction}
\label{sec:mech-stage1-kmer-extraction}

\begin{figure}[h]
    \centering
    \includegraphics[width=.85\linewidth]{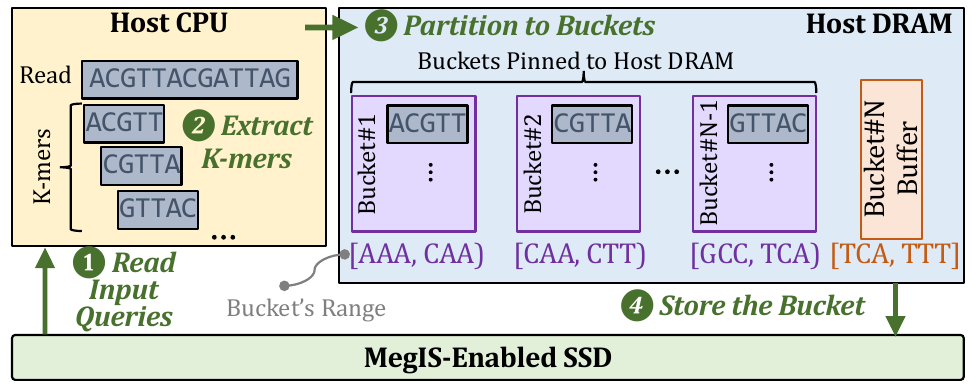}
    \caption{\omii{Overview of the} k-mer extraction \omii{process in \proposal}.}
    \label{fig:kmer-gen}
\end{figure}

To reduce data transfer overhead between \omi{\omii{different parts of the} application that execute in the host system and in the storage system}, we propose a new input processing scheme by improving upon \hhl{the} input processing \omi{scheme} in KMC~\cite{kokot2017kmc3}. We partition the k-mers into buckets, each corresponding to a lexicographical range. 
This enables overlapping the k-mer sorting and transfer of a bucket \omi{to the SSD} with \omi{the} ISP \omi{operations of Step 2 (\sect{\ref{sec:mech-stage2}})} on previous\omii{ly transferred} buckets. \omi{This is}
\bback{because the database k-mers are also sorted and can already be accessed within the 
corresponding range}.
\fig{\ref{fig:kmer-gen}} shows an overview \omi{of \proposals k-mer extraction}. 
The host reads the input reads \omii{from the storage system} (\circled{1} in \fig{\ref{fig:kmer-gen}}), extracts their k-mers (\circled{2}), and stores them in the buckets (\circled{3}).\footnote{\bback{To prevent bucket size imbalance, we initially create preliminary buckets for a small k-mer subset. In case of imbalance, we merge some buckets to \omiii{satisfy} a user-defined bucket count (default 512).}} 
In situations where a sample's extracted k-mers do not fit in the host DRAM, \proposal pins some buckets to \omii{the host} DRAM 
\bback{(e.g., Buckets $1$ to $N-1$ in \fig{\ref{fig:kmer-gen}})} 
and uses the SSD to store the others. This way, k-mers belonging to buckets \omii{in the host DRAM} do \omii{\emph{not}} move back and forth between the host \omii{DRAM} and the SSD (\circled{4}). To reduce the overhead of accessing buckets \omii{in the SSD}, \proposal takes two measures. 
\omii{First, \proposal allocates buffers in the host DRAM specifically for buckets \omii{in the SSD}. Once these buffers are full, it efficiently transfers their contents to the SSD, maximizing the use of the sequential-write bandwidth}. Second, we map \omii{each} bucket's k-mers across SSD channels evenly for parallelism.
\bback{Since \proposal does not require writes to the flash chips after this step \omii{(\omiii{i.e.}, K-mer Extraction in Step 1)}, it can flush all of the FTL metadata for write-related management to free up internal DRAM for the next steps (details in \sect{\ref{sec:mech-ftl}})}.

\subsubsection{Sorting}
\label{sec:mech-stage1-sorting}
\hm{After generating all k-mer buckets,}
\proposal proceeds to sort the k-mers within the individual buckets. As soon as a specific bucket $i$ is sorted, \proposal transfers this bucket to the DRAM inside the SSD in batches (to undergo \omi{Step 2}, as described in \sect{\ref{sec:mech-stage2}}). 
Meanwhile, during the transfer of bucket $i$, \proposal advances to sort bucket $i+1$. 
 \proposal can orthogonally use a sorting accelerator \omii{(e.g.,~\cite{samardzic2020bonsai,qiao2022topsort,jayaraman2022hypersort})} to perform sorting.

\subsubsection{Excluding K-mers}
\label{sec:kmer-exclusion}
\proposal, like various tools\omii{~\cite{benoit2016multiple,bovee2018finch,lapierre2020metalign,ounit2015clark,li2018minimap2,kokot2017kmc3}}, can exclude k-mers based on user-defined frequencies to improve accuracy. Users can exclude 1)~overly common \revh{(i.e., indiscriminative)} k-mers
and 2)~\omii{very infrequent} k-mers \omii{(e.g., those} that appear only once\omii{)}, which may represent 
\bback{sequencing} 
errors or low-abundance organisms \bback{that are hard to distinguish from random occurrences}.
\bback{Exclusion follows sorting, where k-mers are \omii{already} counted}. 
While the size of the extracted query k-mers (\sect{\ref{sec:mech-stage1-kmer-extraction}}) can be large (on average 60~GB in our experiments),
the size of the k-mer set selected to go to \omi{Step 2} is much smaller (on average 6.5~GB) \bback{and is significantly smaller than the database that may reach several terabytes\omii{~\cite{ncbi2023,karasikov2020metagraph,shiryev2023indexing,pebblescout,lemane2023kmindex,marchet2023scalable}}}.

\subsection{Step 2: Finding Candidate \rev{Species}}
\label{sec:mech-stage2}

In \omii{Step 2}, \proposal finds the \hm{species} present in the sample \bback{by 1) intersecting the query k-mers and the database \omi{k-mers},
and 2) finding the \omii{taxID}s of the intersecting k-mers}.
We perform this stage inside the SSD since it requires streaming the large database with low reuse and involves only lightweight computation. 
This enables \proposal to leverage the \omi{SSD's} large internal bandwidth 
 and alleviate the 
 \bback{overall}
 burden of moving\bback{/analyzing}
 large, low-reuse data from the rest of the system.

\bback{Considering the SSD's hardware limitations, \proposal should leverage the full internal bandwidth \omii{without} requir\omii{ing} expensive hardware \omi{resources} inside the SSD (e.g., large \omi{internal} DRAM \omii{size/}bandwidth and costly logic units). Performing this step effectively inside the SSD requires efficient coordination between the SSD and the host, mapping, hardware design, and storage technology-aware algorithmic optimizations}.

\subsubsection{Intersection Finding}
\label{sec:mech-stage2-1}

In this step, \proposal finds the intersecting k-mers, i.e., k-mers present in both the query k-mer buckets arriving from the host \omi{system}
and the large k-mer database stored in the flash chips.

Relying \omii{solely} on the \omii{SSD's} internal DRAM \revh{to} 1) buffer the query k-mers arriving from the host \omi{system} and the database k-mers arriving from the SSD channels at full bandwidth and 2) stream through both to find their intersection \omii{can pressure the valuable internal DRAM} bandwidth.
For example, reading \omii{the database from the SSD channels} at full bandwidth in a high-end SSD can already exceed the LPDDR4 DRAM bandwidth used in current SSDs~\cite{zou2022assasin, samsung980pro,samsungPM1735} and even the \mbox{16-GB/s} DDR4 bandwidth~\cite{zou2022assasin,ddr4sheet}. To address this challenge, we adopt an approach similar to \cite{zou2022assasin} and 
\omii{operate on data fetched from flash chips without buffering them in the internal DRAM}.
Despite its benefits, this approach requires large buffers (64 KB for input and 64 KB for output) \emph{per channel}. 

To facilitate low-cost computation on flash data streams, we leverage two key features of \proposal to find the minimum required buffer size. First, the computation in this step is lightweight and does not require a large buffer for data awaiting computation. Second, data is uniformly spread across channels, with each compute unit handling data from one channel. 
Based on these, we directly read data from the flash \omii{chips} and include \emph{two} k-mer registers per channel. 
One register holds a k-mer as the computation input, 
\omii{while the other register \omiii{stores} the subsequent k-mer as it is read from the flash chips}.
This way, by only using two registers, \proposal directly compute\omii{s} on the flash data stream at low cost. 

\begin{figure}[t]
    \centering
    \includegraphics[width=0.85\linewidth]{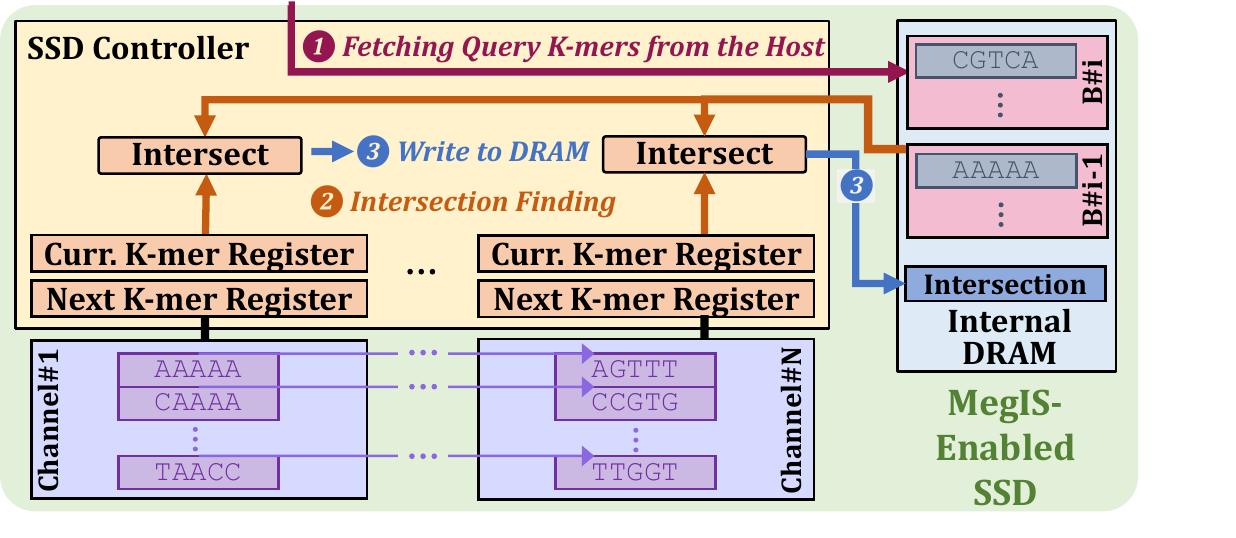}
    \caption{\omii{Overview of the} intersection finding \omii{process in \proposal}.}
    \label{fig:intersection-finding}
\end{figure}

\fig{\ref{fig:intersection-finding}} shows the \omi{overview of \proposals intersection finding \omiii{process}}.
First, \proposal reads the query k-mers to the internal DRAM in batches (\circled{1} in \fig{\ref{fig:intersection-finding}}). Second, it concurrently reads both the sorted query k-mers \omii{(from the internal DRAM)}
and the sorted database k-mers \omii{(from the flash chips)},
performing a comparison to find their intersection (\circled{2}) using per-channel \emph{Intersect} units located \omii{in} the SSD controller.
Third, \proposal writes the intersecting k-mers to the internal DRAM for further analysis (\circled{3}).

\head{Fetching Query K-mers}
To efficiently use external bandwidth, \proposal moves buckets from the host \omi{system} to the internal DRAM in batches.
We manage two batches in the internal DRAM to overlap transfer and intersection finding.
For an SSD with 8 channels, 4 dies/channel, 2 planes/die, and 16-KiB pages, \proposal requires space for two 1-MiB batches\revid{\label{rev:A5.1}A5.1} \irev{(i.e., $B\#i-1$ and $B\#i$ in \fig{\ref{fig:intersection-finding}})} in the internal DRAM.

\head{Intersection Finding}
\proposal reads the query k-mers from the internal DRAM and the database k-mers from the flash \omii{chips}. \omii{Intersection Finding} runs in a pipeline\omi{d manner} with \omii{Fetching Query K-mers}. 
\bback{We store the database evenly across different channels to leverage the full internal bandwidth when sequentially reading data using multi-plane operations.  \proposal finds the intersecting k-mers as follows:} 
If a database k-mer equals a query k-mer, \omi{\proposal} records \omii{the k-mer} as an \omii{intersecting k-mer}. If a query k-mer is larger (smaller), \proposal reads the next database (query) k-mer. 
{\proposal}’s Control Unit, located on the SSD controller, receives the comparison results and issues the control signals accordingly.\footnote{\omii{\figs{\ref{fig:kmer-gen}, \ref{fig:intersection-finding}, and \ref{fig:taxid-finding}}} exclude Control Unit and its connections for readability.}

\head{Storing the Intersecting K-mers}
\proposal stores the intersecting k-mers in the \omii{SSD's} internal DRAM.\footnote{The \omii{intersecting k-mers do} not have a strict size requirement and can use the \irev{available} internal DRAM's space opportunistically. Usually, its small size allows it to fully fit in the internal DRAM. However, in a case where it does not, \proposal starts the \omii{taxID} retrieval (\sect{\ref{sec:mech-stage2-2}}) for the already-found \omii{intersecting k-mers}; then resumes this step, overwriting the old \omii{intersecting k-mers}.} 
The internal DRAM needs to support 1) fetching the queries, 2) reading them out, 3) storing the intersection, and 4) \omii{reading} FTL metadata. Since the query k-mer set, the intersection, and the FTL metadata (FTL details in \sect{\ref{sec:mech-ftl}}) are significantly smaller than the database, they can be accessed at a much smaller bandwidth than \omii{the bandwidth required for} reading the database. 
For example, for our datasets in \sect{\ref{sec:methodology}}, when fully leveraging \ssdp's internal \omii{flash} bandwidth \omii{by reading the database from all flash channels}, \proposal requires only 2.4~GB/s of DRAM bandwidth \omi{to access all datasets stored in the internal DRAM}.

\subsubsection{Retrieving \omiii{TaxID}s}
\label{sec:mech-stage2-2}

\omiii{\proposal finds the \omiii{taxID}s of the species corresponding to the intersecting k-mers by looking up the intersecting k-mers in a pre-built sketch database.
Each sketch is a small representative subset of k-mers associated with a given \omiii{taxID}. A sketch database stores the k-mer sketches and their associated \omiii{taxID}s for a given set of species}.
Similar to \cite{lapierre2020metalign}, we use CMash~\cite{liu2022cmash} to generate sketches. \proposal can \omiii{also} use other sketch \omiii{generation methods}. \proposal flexibly supports variable-sized k-mers \omii{in its sketch database}.  
As shown by prior works~\cite{liu2022cmash,kim2016centrifuge}, while longer k-mers \omiii{are more unique and} offer greater discrimination, they may result in missing matches between \omiii{the intersecting k-mers} and sketches. In such cases, users may also \omiii{search for} smaller k-mers \omiii{by looking up the prefixes of the intersecting k-mers in the sketch database}.
\omiii{This enables finding} additional matches \omiii{and increasing the true positive rate}.

\begin{figure}[b]
    \centering
    \includegraphics[width=0.9\linewidth]{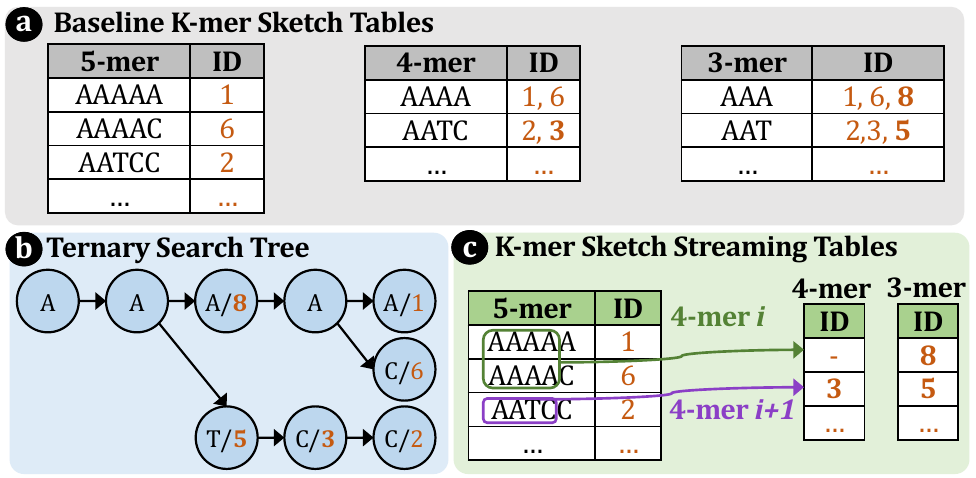}
    \caption{\omii{Overview of} sketch data structures.}
    \label{fig:taxid-structures}
\end{figure}

Finding \omiii{taxID}s \omiii{for} variable-sized k-mers is challenging since it requires many pointer-chasing operations on a large data structure that may not fit in the \omiii{SSD's} internal DRAM.
To \omiii{support variable-sized k-mers}, some approaches \omiii{(e.g.,~\cite{marchet2021data,liu2022cmash,lapierre2020metalign,kim2016centrifuge,song2024centrifuger}}\omiii{)} \omiii{provide} data structures to encode the k-mer information in a space-efficient manner.
For example, CMash~\cite{liu2022cmash} encodes k-mers \gram{of} variable 
sizes in a ternary search tree. \fig{\ref{fig:taxid-structures}} shows \omiii{sketch databases with variable-sized k-mers (k = 5, 4, and 3)} along\rev{side} 
their \omiii{taxID}s in \circled{a} separate tables, as used by some prior approaches~\cite{koslicki2016metapalette,weging2021taxonomic}, and \circled{b} in a ternary search tree. 
This tree structure is devised to 1) save space and 2) retrieve the \omiii{taxID}s for all k-mers with $k \leq k_{max}$ that are prefixes of \hm{a query} $k_{max}$-mer. 
\omiii{For example, as shown in \fig{\ref{fig:taxid-structures}}, when traversing the tree to look up the 5-mer \texttt{AATCC}, we can look up the 4-mer \texttt{AATC} during the same traversal}. 
Despite its benefits, this approach requires up to $k_{max}$ pointer-chasing \gram{operations} for \emph{each lookup}. Performing these operations inside the SSD is challenging since the tree can be larger than the \omiii{SSD's} internal DRAM, and pointer chasing on flash arrays is expensive due to their significantly larger latency compared to DRAM.

While \proposal can \omii{perform \omiii{taxID} retrieval} in the host \omiii{system}, we identify a new optimization opportunity leveraging unique features of ISP (i.e., large internal bandwidth and storage capacity),
which avoids pointer-chasing at the cost of larger data structures. 
\fig{\ref{fig:taxid-structures}} \circled{c} shows an overview of our approach, \emph{K-mer Sketch Streaming (KSS)}. 
For k-mers with \omii{k =} $k_{max}$, \proposal stores the k-mer sketches and their \omiii{taxID}s similar to \circled{a}. \proposal\omii{~keeps} this table \omiii{in a lexicographically-sorted order}.
For each smaller k-mer \omiii{(with k $< k_{max}$)}, \omiii{\proposal} only stores the \omiii{taxID}s that are \emph{not} attributed to their corresponding larger, more unique, k-mer. 
\omiii{For these smaller k-mers, \proposal does not store the k-mer itself and instead, uses the prefixes of the \mbox{$k_{max}$-mers} to retrieve the smaller k-mer\omiii{s}}.
\omiii{\proposal} allows for \omiii{taxID} retrieval by sequentially streaming through the intersecting k-mers  
\bback{(which are already sorted)}
and the KSS tables. 
While \gram{the} KSS data structure is larger than \omiii{the corresponding ternary search tree} \circled{b}, it is \omiii{much more} suitable for ISP due to \omiii{its streaming access feature}. KSS can also be \omiii{efficient} for \omiii{processing} outside \omiii{the} storage \omiii{system} with SSDs with \hm{high} external bandwidth (\sect{\ref{sec:eval-main}}). KSS leads to 7.5$\times$ smaller data structures compared to the 107-GB \omi{data structure in} \circled{a}, and 2.1$\times$ larger compared to \circled{b} (dataset details in \sect{\ref{sec:methodology}}).

\begin{figure}[h]
    \centering
    \includegraphics[width=0.9\linewidth]{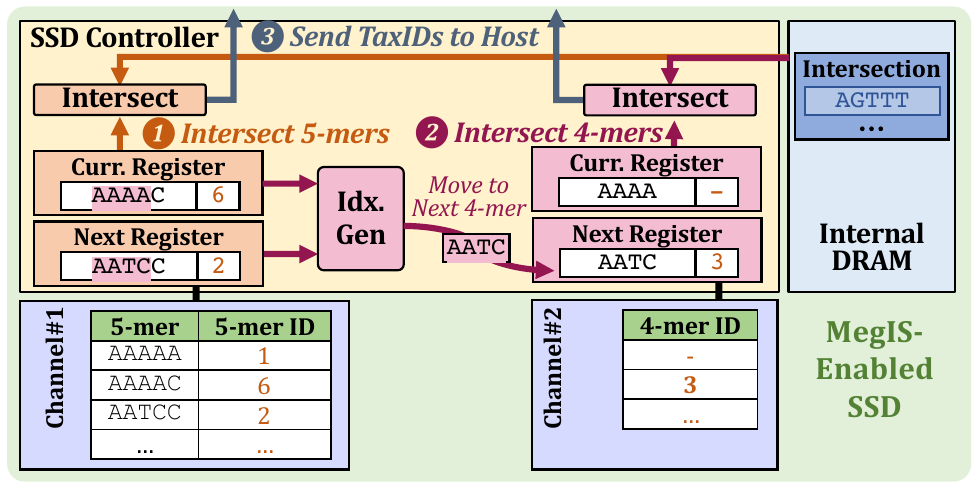}
    \caption{\omiii{Overview of the TaxID} retrieval \omiii{process in \proposal}.}
    \label{fig:taxid-finding}
\end{figure}

\fig{\ref{fig:taxid-finding}} shows \gram{an} overview of \proposal's \omiii{taxID} retrieval \omiii{process}. As an example, we demonstrate retrieving 5- and 4-mers. First, \proposal reads the intersecting k-mers (i.e., 5-mers) from the internal DRAM and concurrently reads the 5-mer sketches and their IDs from an SSD channel to find their matches (using the same Intersect unit in \sect{\ref{sec:mech-stage2-1}})  (\circled{1}). 
Second, to find 4-mer matches, \proposal compares the \emph{prefixes} of the intersecting 5-mers with the prefixes of the 5-mer sketches (\circled{2}). 
\proposal~\omiii{incorporates} a lightweight \emph{Index Generator}.
\omiii{It compares the 4-mer prefixes of each pair of consecutive 5-mers. When the prefixes differ (indicating the start of a new 4-mer), it identifies the new prefix as the new 4-mer and reads the next 4-mer \omiii{taxID} from a SSD channel.}
Third, \proposal sends the retrieved \omiii{taxID}s to the host (\circled{3}) as the IDs of the candidate species present in the sample.

\subsection{Step 3: Abundance Estimation}
\label{sec:mech-stage3}

For applications \omiii{that require} abundance estimation, 
\bback{\proposal integrates further analysis on the candidate species identified as present \omiii{in the sample} at the end of Step 2}. 
\proposal can flexibly integrate with different approaches \omiii{to abundance estimation} used in various tools,
such as \rev{\inum{i}}~lightweight statistics \omiii{(e.g.,~\cite{lu2017bracken,dimopoulos2022haystac,koslicki2016metapalette})}
or \inum{ii}~more accurate and costly read mapping \omiii{(e.g.,}~\cite{lapierre2020metalign,kim2016centrifuge,milanese2019microbial}\omiii{)}, \bback{where the input read set is mapped to the reference genomes of candidate species \omiii{present in the sample}}. \omiii{Based on the relative number of reads that map to each species' reference genome, we can determine the occurrence frequencies of different species}.
\irev{\proposal can\revid{\label{rev:C3}C3} integrate with different existing statistical approaches or \omiii{read mapping, performed in the host or an accelerator,} specialized for short reads (e.g.,\hhl{~\cite{cali2020genasm,nag2019gencache,fujiki2018genax}}) or long reads (e.g.,\hhl{~\cite{turakhia2018darwin,cali2020genasm,nag2019gencache}}). \omiii{We note that Steps 1 and 2} of \proposal are based on k-mers extracted from the reads and do not depend on a specific read length}.

\begin{figure}[t]
    \centering
    \includegraphics[width=0.8\linewidth]{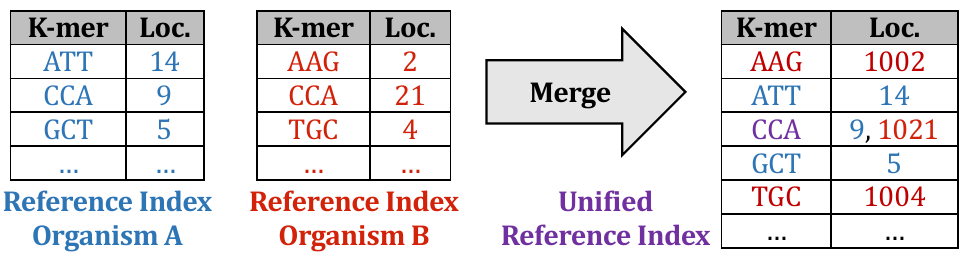}
    \caption{Merging the reference indexes \omiii{to facilitate read mapping during abundance estimation}.}
    \label{fig:index-merger}
\end{figure}

\omiii{While the lightweight statistical approaches can work directly on the output of Step 2, \proposal requires additional data preparation to facilitate read mapping}.
The \omiii{read} mapper \omiii{requires} the query reads and a \omiii{unified} index of the reference genomes of the candidate species \omiii{present in the sample}~\cite{li2018minimap2}. \omiii{In comparison to using individual indexes for each species, the unified index eliminates the need to search through each index separately, thereby reducing the overheads of the \omiii{read} mapping process}.
\bback{Building \omiii{indexes for} individual species is a one-time task. Yet, creating a unified index for the initially unidentified species present in the sample cannot be done offline}.
\omiii{\proposal} facilitates \omiii{index generation for read mapping} by generating a unified 
index 
in the SSD.
\omiii{\fig{\ref{fig:index-merger}} shows an example of the process of unified index generation in \proposal}.
\bback{Each index entry shows a k-mer and its location in that species' reference genome}.
\proposal reads each index \omiii{stored in the flash chips} sequentially and merges \omiii{their entries into a unified index}. 
\bback{When \omiii{\proposal finds} a common k-mer (e.g., \texttt{CCA} in \fig{\ref{fig:index-merger}}), it stores \omiii{the corresponding location of the k-mer in both reference genomes,} adjusting the locations with appropriate offsets based on \omiii{the reference genome sizes}}. \omiii{After generating the unified index, \proposal transfers the index to the host system or an accelerator to perform read mapping for abundance estimation}.

\subsection{\omiii{\proposal{}~FTL}}
\label{sec:mech-ftl}

\omiii{\proposal{}~FTL} needs simple changes to the baseline FTL to handle communication between the host and the SSD.

\omiii{\head{FTL Metadata}} At the beginning \omiii{of \proposals operation as a metagenomic acceleration framework}, \omiii{\proposal{}~FTL} maintains all metadata of the regular FTL in the internal DRAM.
For the only step \omiii{that} require\omiii{s} writes \omiii{to the NAND flash chips} (\sect{\ref{sec:mech-stage1-kmer-extraction}}, \omiii{K-mer Extraction} in the host), \omiii{\proposal{}~FTL} uses the write-related metadata \omiii{(e.g., L2P, bad-block information)}. 
After \omiii{the K-mer Extraction step}, \proposal does not require writes \omiii{to the NAND flash chips}, so it flushes the regular L2P \omiii{metadata} and loads \omiii{\proposal{}~FTL}'s L2P \omiii{metadata} while still keeping the other metadata \omiii{of a regular FTL}.
 
\omiii{\proposal is designed to only access the underlying flash chips \emph{sequentially}, which inherently reduces the size of the required L2P mapping metadata.
In regular FTL, L2P mappings dominate the SSD's internal DRAM capacity due to the page-level granularity of mappings~\cite{mansouri2022genstore, kim-dac-2017, samsung870evo}. However, by accessing data sequentially, \omiii{\proposal{}~FTL} circumvents the need for such detailed page-level mappings. Instead, \omiii{\proposal{}~FTL} utilizes a more coarse-grained block-level mapping, which substantially reduces the size of L2P metadata.
Therefore, flushing regular L2P mapping metadata into flash chips and using \omiii{\proposal{}~FTL}'s metadata enables us to exploit most of the internal DRAM \omiii{bandwidth and capacity} during ISP}.

\begin{figure}[b]
    \centering
    \includegraphics[width=0.9\linewidth]{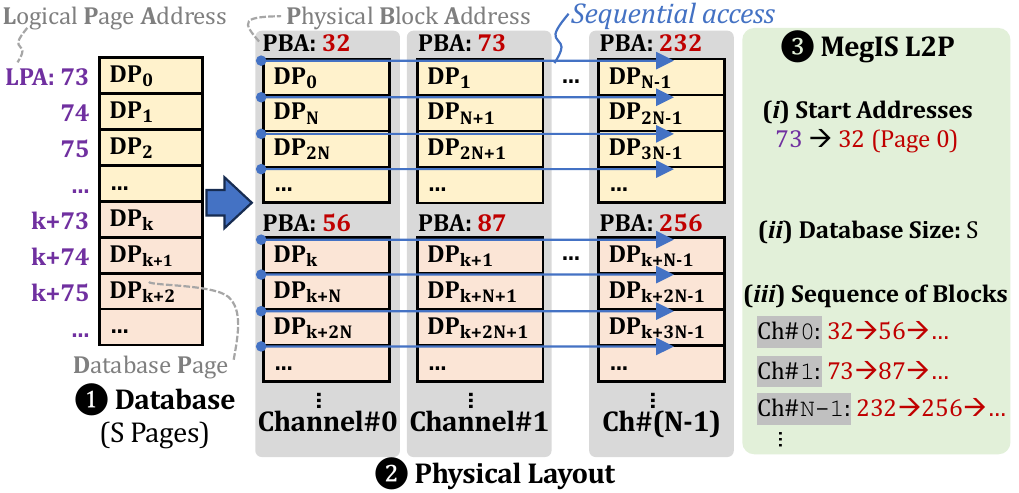}
    \caption{Data layout and mapping data structure in \proposal.}
    \label{fig:ms-ftl}
\end{figure}

\omiii{\head{Data Placement}} \fig{\ref{fig:ms-ftl}} shows how \omiii{\proposal{}~FTL} manages the target data \omiii{(i.e., databases\footnote{K-mer \omiii{databases (\sect{\ref{sec:mech-stage2-1}})} and sketch databases \omiii{(\sect{\ref{sec:mech-stage2-1}})} are the only data structures accessed from NAND flash memory during \proposal's ISP operations.})} stored in NAND flash with reduced \omiii{L2P} metadata.
When storing a database \omiii{in the SSD} \circled{1}, \omiii{\proposal{}~FTL}  \emph{evenly and sequentially} distributes the data across all  channels \omiii{\circled{2}} while ensuring that every \emph{active} block~\cite{kawaguchi1995flash} 
\bback{(\omiii{i.e., blocks available for write operations in the SSD})}
in different channels has the same page offset.
\omiii{Since \proposal always accesses the database \emph{sequentially}, \omiii{\proposal{}~FTL}'s L2P mapping metadata} \circled{3} consists of \inum{i} the mapping between start logical page address \omiii{(LPA)} and physical page address \omiii{(PPA)}, \inum{ii} the database size, and \inum{iii} the sequence of physical block addresses storing the database.
As shown in \fig{\ref{fig:ms-ftl}}, \omiii{\proposal{}~FTL} can sequentially read the stored database \omiii{from the starting LPA}
while performing
reads in a round-robin manner across channels. \omiii{To do so, \omiii{\proposal{}~FTL} just increments the PPA within a physical block and resets the PPA when reading the next block}.

Compared to the regular L2P, whose space overhead is 0.1\% of stored data (4 bytes per 4 KiB), \proposal's L2P is very small. 
For example, \proposal only requires $\sim$1.3 MB to store a 4-TB database, assuming a physical block size of 12~MB: 4~bytes for each of the 349,525
used blocks (and a few bytes for the start L2P mapping and database size).
The only metadata other than L2P that must be kept during ISP is the per-block access count for read-disturbance management~\cite{cai2015read}, so the total \proposal\irev{-FTL}\revid{\label{rev:A5.3}A5.3} metadata size is up to 2.6~MB.

\omiii{\head{SSD Management Tasks}} \proposal's ISP accelerators are located \omiii{in} the SSD controller and access data \emph{after} ECC. \irev{ECC\revid{\label{rev:C2}C2} does not restrict \proposals ISP performance. Modern SSDs are designed with ECC capabilities that match the full internal bandwidth \omiii{of the SSD} to support both I/O requests and internal data migrations due to management tasks like garbage collection\omiii{~\cite{kim2023decoupled,cai-insidessd-2018,cai2017error}}. 
\icut{Otherwise, ECC would limit the SSD’s performance, which would an unreasonable design. In many cases, modern SSDs incorporate per-channel ECC that is shared between multiple chips connected to a channel\hhl{\omiii{~\cite{kim2023decoupled,cai-insidessd-2018,cai2017error}}}. Even where multiple channels share an ECC, the ECC still needs to match the full internal bandwidth   to avoid bottlenecking the SSD’s performance.}}

\proposal performs other tasks for ensuring reliability (e.g., refresh to prevent uncorrectable errors\cite{cai-hpca-2017, luo2018improving,luo-hpca-2018,cai2015read,cai2013error, cai2012flash, cai2017error, ha2015integrated, cai-insidessd-2018,luo2015warm}) \emph{before or after} the ISP because 1) the duration of each \proposal process is significantly smaller than the manufacturer-specified threshold for reliable retention age (e.g., one year~\cite{micron3dnandflyer}), and 2) \proposal avoid\hhl{s} read disturbance errors~\cite{cai2015read} during ISP due to its sequential low-reuse accesses.

\subsection{\irev{Storage Interface Commands\revid{\label{rev:A7}A7}}}
\label{sec:mech-interface}

\irev{\proposal requires three new NVMe commands. First, \textsf{MegIS\_Init} initiates the metagenomic analysis and communicates the size and starting address of the space in the host DRAM that is available for \proposals operations. \omiii{Upon receiving this command,} \proposal readies itself to work in the metagenomic acceleration mode \omiii{(\sect{\ref{sec:mech-overview}})}. During metagenomic analysis steps, \omiii{\proposal{}~FTL} and \proposal's FSM controller handle the data/control flow. 
Second, \textsf{MegIS\_Step} communicates the start and end of \omiii{each} step executed in the host to the SSD, enabling \proposal to manage control and data flow accordingly.  \omiii{\textsf{MegIS\_Step} specifies the step performed in the host system, such as k-mer extraction (\sect{\ref{sec:mech-stage1-kmer-extraction}}) or sorting (\sect{\ref{sec:mech-stage1-sorting}}), with an input argument. Each time this command with the same argument is sent, it alternates between marking the start and the end of a step}.
After completing the metagenomic analysis, \proposal switches back to operating as a baseline SSD. Third, \textsf{MegIS\_Write} is a specialized write operation that updates \omiii{\proposal{}~FTL}'s small mapping metadata whenever metagenomic data is written to the SSD}. \omiii{\textsf{MegIS\_Write} is similar to the regular NVMe write command, except that it updates mapping metadata in both the regular FTL and \omiii{\proposal{}~FTL}}.

\subsection{Multi-Sample Analysis}
\label{sec:mech-multi-sample}

\newcommand\bases{\textsf{Base-S}\xspace}
\newcommand\basem{\textsf{Base-M}\xspace}
\newcommand\mss{\textsf{MS-S}\xspace}
\newcommand\msm{\textsf{MS-M}\xspace}
\newcommand\optm{\textsf{Opt-M}\xspace}

\begin{figure}[h]
    \centering
    \includegraphics[width=0.9\linewidth]{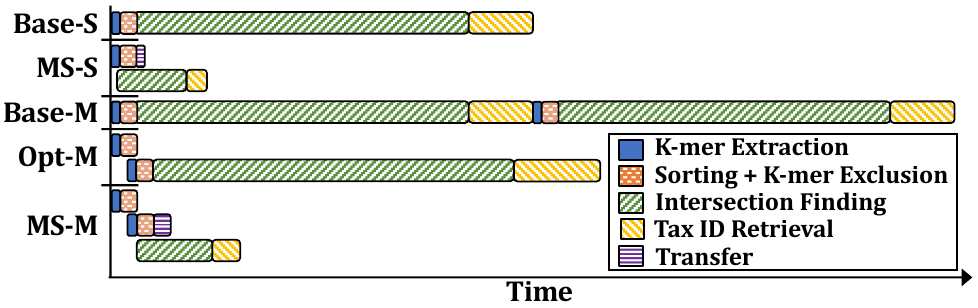}
    \caption{\new{Timeline of \omiii{analyzing single or multiple samples in the baseline and \proposal.}}}
    \label{fig:multi-timeline}
\end{figure}

For some use cases \omiii{(e.g., globally tracing antimicrobial resistance\cite{danko2021global}, associating gut microbiomes to health status~\cite{Turnbaugh2007,hhrlich2011metahit})}, 
a metagenomic study can have multiple read sets (i.e., samples) that need to access the \omiii{same} database.
If the host's DRAM is larger than the k-mer sizes extracted from a sample, we use the available DRAM opportunistically to buffer k-mers extracted from \emph{several} samples. This way, \proposal streams through \omiii{one} database only once.
\fig{\ref{fig:multi-timeline}} shows the timeline of analyzing \gram{a} single (\textsf{S}) or multiple (\textsf{M}) samples in the baseline (\textsf{Base}), in our proposed optimized approach in software (\textsf{Opt}), and in \proposal (\textsf{MS}).
\omiii{To accelerate input query processing (\sect{\ref{sec:mech-stage1}}) when analyzing several input query samples,} \proposal can \omiii{be} flexibly integrate\omiii{d} with a sorting accelerator \omiii{(e.g.,~\cite{samardzic2020bonsai,qiao2022topsort,jayaraman2022hypersort})} \omiii{and} further improve end-to-end performance.

\section{Evaluation\revid{\label{rev:B3}B3} Methodology}
\label{sec:methodology}

\head{Performance} 
We design a simulator that models all \hhl{of} \proposal's components, including host operations, accessing flash chips, internal DRAM, in-storage accelerator, and host-SSD interfaces. We feed the latency and throughput of each component to this simulator.  
For the components in the \textbf{hardware-based} steps (e.g., ISP units in Steps 2 and 3):  We implement \proposal's logic components in Verilog. We synthesize them using the Synopsys Design Compiler~\cite{synopsysdc} with a 65~nm library\cite{umcL65nm} and \irev{perform place-and-route using Cadence Innovus\hhl{~\cite{innovus}}}.
We use two state-of-the-art 
simulators, Ramulator~\cite{kim2016ramulator, ramulatorsource} to model SSD's internal DRAM, and MQSim\omii{~\cite{tavakkol2018mqsim,mqsimsource}} \omii{to model} SSD's internal operations.
For the components in the \textbf{software-based} step (e.g., host operations in Step 1)\omii{, we} measure performance on a real system, an AMD$^\text{\textregistered}$ EPYC$^\text{\textregistered}$ 7742 CPU\omii{~\cite{amdepyc}} with 128 physical cores \hhl{and} 1-TB DRAM (in all experiments unless stated otherwise). 
For the software baselines, we measure performance on this real system, with best-performing thread counts.
\omii{The source code of \proposal, scripts, and datasets can be freely downloaded from https://github.com/CMU-SAFARI/MegIS.}

\head{SSDs} We use \ssdc~\cite{samsung870evo} and \ssdp~\cite{samsungPM1735} as described in \sect{\ref{sec:motivation-ovhd}} \irev{in our real system experiments. In our MQSim simulations for the ISP steps, we faithfully model the SSDs with the configurations summarized in \revid{\label{rev:A3}A3}Table~\ref{table:SSD_config}}.

\begin{table}[t]
\vspace{5pt}
\centering
\scriptsize
\color{black}
\caption{\irev{SSD configurations.}}
\begin{tabular}{@{\hspace{-0.5pt}}c@{\hspace{-0.05pt}}|c|c@{\hspace{-0.05pt}}}
\toprule
\textbf{Specification} & \textbf{SSD-C} & \textbf{SSD-P} \\ 
\midrule
\midrule
\textbf{General}       & \multicolumn{2}{c}{\begin{tabular}[c]{@{}c@{}} 48-WL-layer 3D TLC NAND flash-based SSD\\ 4 TB capacity, 4 GB internal \omii{LPDDR4} DRAM\omii{~\cite{lpddr4}} \end{tabular}} \\ \midrule
\begin{tabular}[c]{@{}c@{}} \textbf{Bandwidth}\\ \textbf{(BW)} \end{tabular}  & \begin{tabular}[c]{@{}c@{}}600 MB/s interface  BW\\ (SATA3);\\ 560 MB/s sequential-read BW\\  1.2-GB/s channel I/O rate\end{tabular} & \begin{tabular}[c]{@{}c@{}}8 GB/s interface BW \\ (4-lane PCIe Gen4);\\ 7 GB/s sequential-read BW\\1.2-GB/s channel I/O rate\end{tabular} \\ \midrule
\begin{tabular}[c]{@{}c@{}} \textbf{NAND}\\ \textbf{Config} \end{tabular}  & 
\begin{tabular}[c]{@{}c@{}} 8 channels, 8 dies/channel,\\ 4 planes/dies, 2,048 blocks/plane,\\ 196  WLs/block, 16 KiB/page \\ \textit{(4/8/16 channels in \fig{\ref{fig:internal-bw-swp}})}\end{tabular}  &
\begin{tabular}[c]{@{}c@{}} 16 channels, 8 dies/channel,\\ 2 planes/dies, 2,048 blocks/plane,\\ 196 WLs/block, 16 KiB/page \\ \textit{(8/16/32 channels in \fig{\ref{fig:internal-bw-swp}})}\end{tabular}  \\ \midrule
\textbf{Latencies}     & \multicolumn{2}{c}{\begin{tabular}[c]{@{}c@{}} \omii{Read (tR): 52.5 $\mu$s,  Program (tPROG): 700 $\mu$s}\end{tabular}} \\ \midrule
\begin{tabular}[c]{@{}c@{}} \textbf{Embedded}\\ \textbf{Cores} \end{tabular}  & 3 ARM Cortex-R4 cores\omii{~\cite{cortexr4}}  & 4 ARM Cortex-R4 cores\omii{~\cite{cortexr4}}  \\ \midrule
\bottomrule
\end{tabular}
\label{table:SSD_config}
\end{table}

\head{Area and Power} 
For logic components, we use the results from our Design Compiler synthesis. 
For SSD power, we use the values of a Samsung 3D NAND flash-based SSD~\cite{samsung860pro}. For DRAM power, we base the values on a DDR4 model~\cite{ddr4sheet, ghose2019demystifying}. For the CPU cores, we use AMD$^\text{\textregistered}$ \textmu{}Prof~\cite{microprof}.

\newcommand\popt{\textsf{P-Opt}\xspace}
\newcommand\aopt{\textsf{A-Opt}\xspace}
\head{Baseline Metagenomic Tools} We use a state-of-the-art performance-optimized (\textsf{P-Opt}) tool, Kraken2 + Bracken~\cite{wood2019improved}, and a state-of-the-art accuracy-optimized (\textsf{A-Opt}) tool, Metalign~\cite{lapierre2020metalign}. 
Particularly, for the presence/absence task, we use Kraken2 without Bracken, and Metalign without mapping \bback{(i.e., only KMC~\cite{kokot2017kmc3} + CMash~\cite{liu2022cmash})}. 
For abundance estimation, we use Kraken2 + Bracken, and full Metalign.
\omii{\aopt achieves significantly \omiii{higher} accuracy compared to \popt~\cite{meyer2021critical,lapierre2020metalign}. In particular, \aopt leads to \omiii{4.6--5.2}$\times$ higher F1 scores and \omiii{3--24}\% lower L1 norm error across \omiii{all tested} inputs. One major reason is that \aopt uses larger and richer databases compared to performance-optimized \popt.  
\proposals end-to-end accuracy matches the accuracy of \aopt because \proposal's databases encode the same set of k-mers and sketches as \aopt.}

For both Metalign and \proposal, we use GenCache~\cite{nag2019gencache} for mapping. 
\bback{We use the mapping throughput as reported by the original paper~\cite{nag2019gencache}.
 \proposal can \omii{be} flexibly integrate\omii{d} with other mappers}. 
We also \irev{evaluate}
a state-of-the-art PIM k-mer matching accelerator~\cite{wu2021sieve} \hm{for} accelerat\hm{ing} Kraken2's pipeline. We use the k-mer matching performance as reported by the original paper~\cite{wu2021sieve}.

\head{Datasets} 
We use three query read sets from the commonly-used CAMI benchmark~\cite{sczyrba2017critical}, \bback{with low, medium, and high \hm{genetic} diversity} \omii{(i.e., CAMI-L, CAMI-M, and CAMI-H, respectively)}.
Each read set has 100 million reads.
We generate a database based on \hm{microbial genomes drawn from NCBI's databases}~\cite{ncbi2020,lapierre2020metalign}\irev{\revid{\label{rev:B5.2}B5.2} including 155,442 genomes for 52,961 microbial species.\icut{ (as listed in\hhl{~\cite{metalign_db}}).} For database generation, we use}  default parameters for each tool. For Kraken2~\cite{wood2019improved}, this results in a 293~GB database. For Metalign~\cite{lapierre2020metalign}, this results in a 701~GB k-mer database
and \gram{a} 6.9~GB sketch tree.
\proposal uses the same 701~GB k-mer database and a 14~GB sketch database for \omii{\proposals} \cmashopt~\omii{sketch database (\sect{\ref{sec:mech-stage2-2}})}.

\section{Evaluation}
\label{sec:eval}

\subsection{Presence/Absence Identification Analysis}
\label{sec:eval-main}

\renewcommand\popt{\textsf{P-Opt}\xspace}
\renewcommand\aopt{\textsf{A-Opt}\xspace}
\newcommand\msnol{\textsf{MS-NOL}\xspace}
\newcommand\msext{\textsf{Ext-MS}\xspace}
\newcommand\mscc{\textsf{MS-CC}\xspace}
\newcommand\msfull{\textsf{MS}\xspace}

\bback{We use 1\gram{-}TB \omii{host} DRAM in this analysis (smaller than all datasets \omii{we evaluate})}. 
\omii{We examine seven metagenomic analysis configurations:}
1)~\popt,
2)~\aopt,
3)~{\aopt}+\cmashopt\omii{, where} \aopt leverages the software implementation of \omii{\proposals} \cmashopt approach (\sect{\ref{sec:mech-stage2-2}}) instead of Metalign's CMash~\cite{lapierre2020metalign} \omii{for retrieving taxIDs}, 
\rev{\omii{4})~\msext: a \proposal implementation
without ISP, where the same accelerators used in \proposal are  outside the SSD},
\omii{5})~\msnol: a \proposal implementation \omii{\emph{without}} overlapping the host and SSD operations as enabled by \proposal's bucketing (\sect{\ref{sec:mech-stage1}}),  
\rev{6)~\mscc: a \proposal configuration
\omii{where} the SSD cores perform \proposal's ISP tasks, and}
7)~\msfull: a \underline{M}egI\underline{S}~\rev{configuration \omii{where} the \omii{hardware} accelerators on the SSD controller perform the ISP tasks}.

\begin{figure}[h]
\centering
\vspace{0.2em}
 \includegraphics[width=0.97\linewidth]{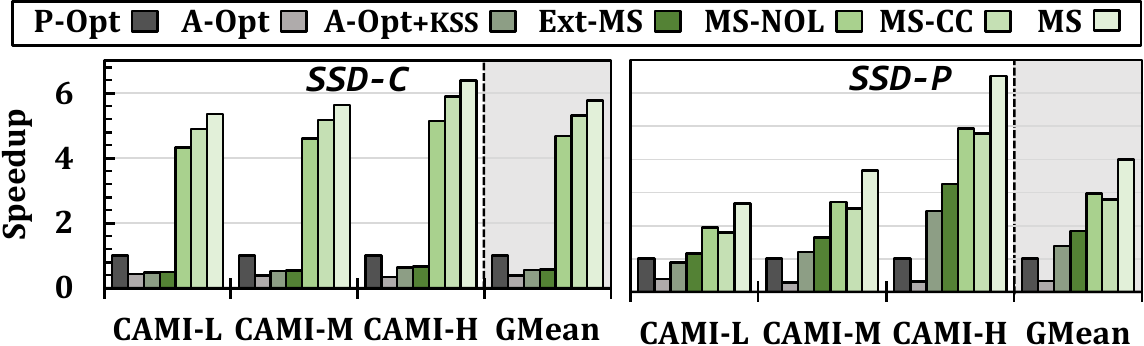}
\caption{\rev{Speedup for different SSDs and input sets.}}
\label{fig:main-eval}
\end{figure}

\omii{\fig{\ref{fig:main-eval}} shows the speedup of the seven configurations over \popt, on three read sets and with \ssdc and \ssdp}. 
\bback{We make six key observations}. 
First, \omii{\proposals full implementation (\msfull) achieves significant speedup compared to both performance-optimized (\popt) and accuracy-optimized (\aopt) baselines}. 
\omii{W}ith \ssdc (\ssdp), \msfull is 5.3--6.4$\times$ (2.7--6.5$\times$) faster compared to \popt, and 12.4--18.2$\times$ (6.9--20.4$\times$) faster compared to \aopt.
Second, \omii{{\aopt}+\cmashopt, which leverages \proposals taxID retrieval approach (\cmashopt) instead of {\aopt}'s baseline taxID retrieval approach,} improves \aopt's performance by 1.4$\times$ (4.2$\times$) on average on \ssdc (\ssdp). \omii{\proposals full implementation outperforms \aopt{}+\cmashopt by} 10.5$\times$ (2.9$\times$). This shows that while \omii{\proposals} \cmashopt approach\gram{,} even outside the SSD\gram{,} provides large benefits, \omii{\proposals full implementation} provides significant additional benefits by alleviating I/O overhead. 
Third, with \ssdc (\ssdp), \msfull leads to 23.5\% (34.9\%) \hm{greater} average speedup compared to \omiii{\proposals implementation without overlapping the steps (}\msnol\omiii{). This is} due to {\proposal}'s bucketing \omiii{scheme} that enables overlapping the steps.
\rev{Fourth, \msfull leads to 10.2$\times$ (2.2$\times$)
average speedup on \ssdc (\ssdp) compared to \omii{\proposals implementation outside the SSD} (\msext)  due to \omii{the benefits of \proposals} specialized ISP}.
\rev{Fifth,
while \mscc provides large speedup,
\msfull leads to 9\% (43\%) greater average speedup compared to \mscc on \ssdc (\ssdp). While both \proposal configurations provide large speedup, this shows \omii{that the hardware} accelerators \omii{are useful and their benefits improve} as the internal bandwidth grows.}
\omii{Sixth, \proposals speedup improves as the genetic diversity of the input read sets increases (from CAMI-L to CAMI-H). This is due to the presence of more species in more diverse read sets, which results in a greater number of sketch tree lookups \omiii{in the baseline} taxID \omiii{retrieval approach}. In contrast, \proposals \cmashopt efficiently retrieves all taxIDs in a single pass through the sketch tables}.

\begin{figure}[h]
\centering
 \includegraphics[width=\linewidth]{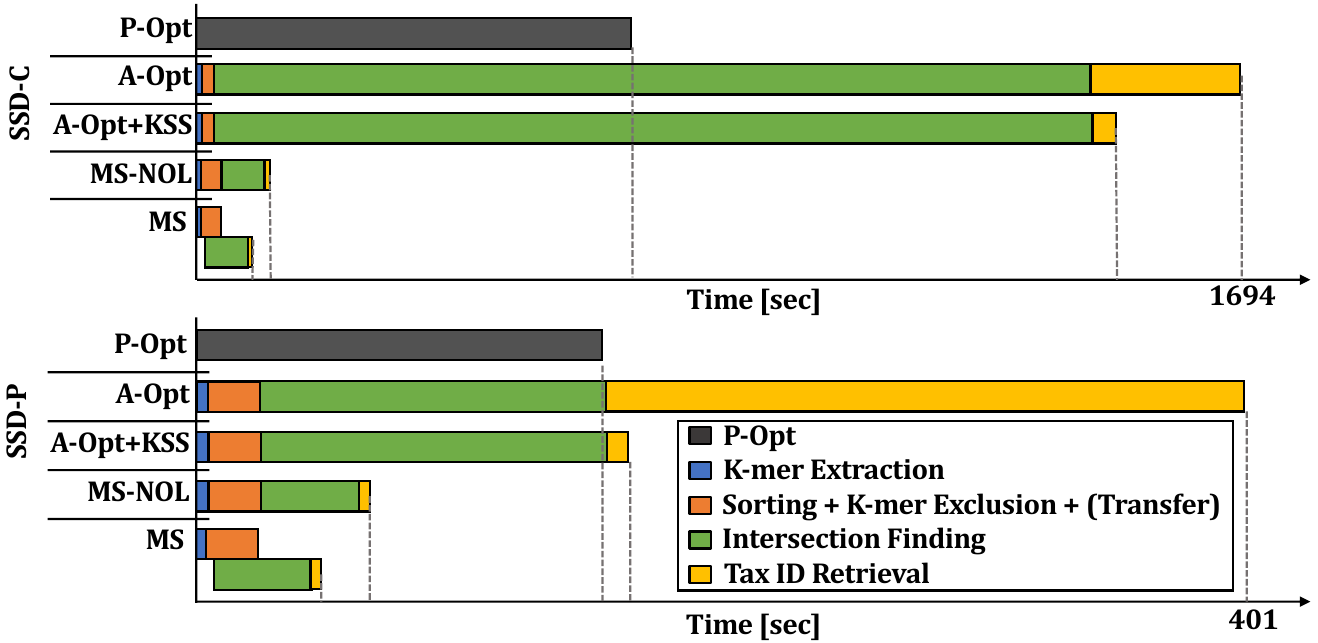}
\caption{\rev{Time breakdown with different SSDs for CAMI-L.}}
\label{fig:exec-breakdown}
\end{figure}

\bback{To further demonstrate the benefits of \proposal's optimizations}, 
\fig{\ref{fig:exec-breakdown}}
shows the time breakdowns with \mbox{CAMI-L} as a representative \omii{input}. 
First, \omii{\cmashopt improves performance by reducing the execution time of taxID retrieval (as seen by \aopt{}+\cmashopt over \aopt).
Second, \proposal without overlapping improves performance over \aopt{}+\cmashopt by leveraging ISP to accelerate intersection finding and taxID retrieval  (as seen by \msnol over \aopt{}+\cmashopt).  
Third, adding overlapping in \proposals full implementation improves performance by overlapping the execution of sorting in the host system with intersection finding in the SSD (as seen by \msfull over \msnol).  
}

\begin{figure}[h]
\centering
 \includegraphics[width=0.9\linewidth]{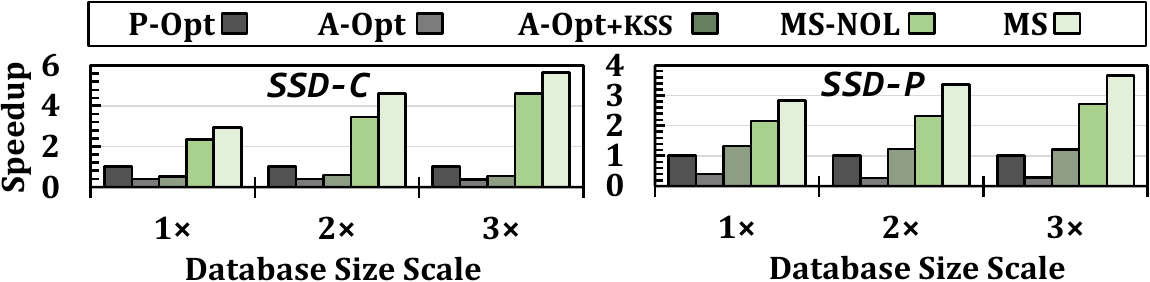}
\caption{\hm{Speedup with} different database sizes.}
\label{fig:eval-db}
\end{figure}

\head{\omii{Effect} of Database Size} \fig{\ref{fig:eval-db}} shows the \omii{effect} of database size, using CAMI-M as a representative input. The largest database size in each tool (marked by 3$\times$) equals the size mentioned in \sect{\ref{sec:methodology}}.
We observe that \proposal's speedups increase as the database size increases (up to 5.6$\times$/3.7$\times$ speedup compared to \popt 
on \ssdc/\ssdp~\omii{as database size grows to 3$\times$}).

\begin{figure}[h]
    \centering
    \includegraphics[width=0.9\linewidth]{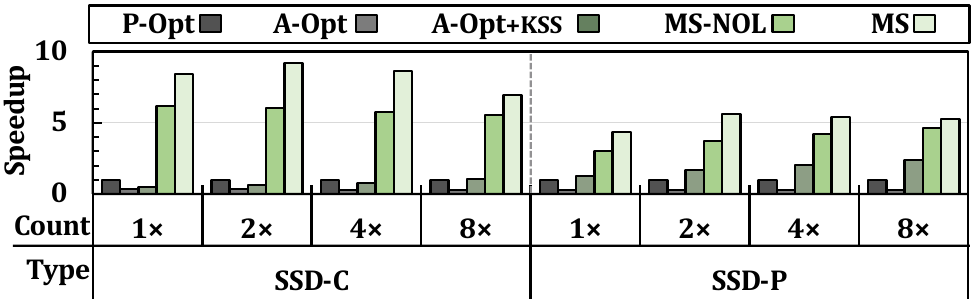}
    \caption{Speedup with different number of SSDs.}
    \label{fig:eval-ssd}
\end{figure}

\head{\omii{Effect} of the Number of SSDs}
\proposal benefits from more SSDs in two ways.
First, mapping different databases to different SSDs allows \hhl{for} concurrent analyses, each benefiting from \proposal, as already shown \omii{(see \fig{\ref{fig:main-eval}})}. \omii{Since} \proposal's databases and queries \omii{are} sorted, the database can be disjointly split across SSDs. \fig{\ref{fig:eval-ssd}} \omii{demonstrates} this case \omii{by showing the speedup of different configurations over \popt}.
We show that \proposal maintains its large speedup with many SSDs \omii{(i.e., up to eight)}. \omii{A}s the external bandwidth increases for the baselines \omii{(with the number of SSDs)}, internal bandwidth also increases for \proposal. Particularly, speedup over \popt increases until some point 
\bback{(two SSDs)}
because \proposal takes better advantage of the scaling due to its more efficient streaming accesses. 
Although there is a slight decrease in speedup when moving from two to eight SSDs, the speedup is still high (6.9$\times$/5.2$\times$ over eight SSD-Cs/SSD-Ps). This decrease is because in \proposal, due to the large internal bandwidth with 8 SSDs, the overall throughput becomes dependent on the host's sorting. Therefore, 
\bback{in systems with many SSDs},
\proposal can \omii{be} integrate\omii{d} with an accelerator for sorting \omii{(e.g.,~\cite{samardzic2020bonsai,qiao2022topsort,jayaraman2022hypersort})} for further speedup. We conclude that \proposal effectively leverages the increased internal bandwidth with more SSDs.
\irev{Due to this efficient use of multiple SSDs \revid{\label{rev:B5.1}B5.1}(owing to \proposals sorted database that can be disjointly partitioned), \proposal can efficiently scale up to very large databases that are distributed \omii{across} different SSDs.}

\begin{figure}[t]
\centering
 \includegraphics[width=0.9\linewidth]{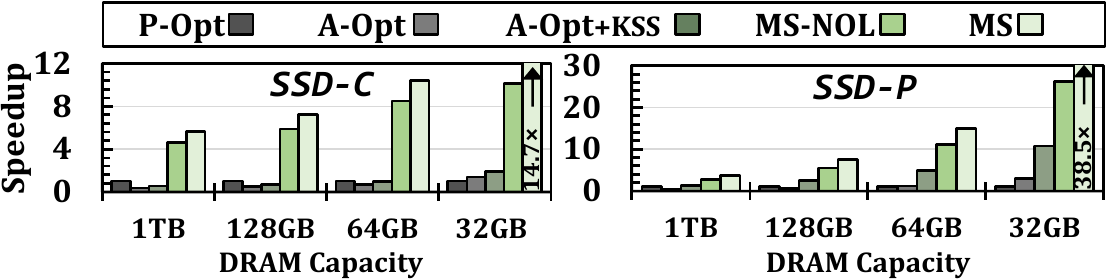}
\caption{\hm{Speedup} with different main memory capacities.}
\label{fig:eval-mem}
\end{figure}

\head{\omii{Effect} of Main Memory Capacity} \fig{\ref{fig:eval-mem}} \omii{demonstrates} the \omii{effect} of \omii{host} DRAM capacity \omii{by showing the speedup of all configurations over \popt}
with CAMI-M.\footnote{In all cases, \rev{except for the 32GB configuration,} all k-mer buckets extracted from the read set (\sect{\ref{sec:mech-stage1-kmer-extraction}}) fit in \omii{the host} DRAM.}
\bback{To gain a fair understanding of I/O overheads when DRAM is smaller than the database, we reduce \omii{I/O} overheads as much as possible in software. We adopt an optimization~\cite{pockrandt2022metagenomic} to load and process \popt's database into chunks that fit in DRAM.\footnote{Note \aopt does not require this due to its streaming database accesses.} In this case, random accesses to the database in each chunk do not repeatedly access the SSD. However, two overheads still remain: 1) there is still the I/O cost of bringing all chunks from the SSD \omii{to the host DRAM}, and 2) for every \omii{database} chunk, all of the input sequences must be queried}.
We make three observations. 
First, \proposal's speedup increases compared to \popt with smaller DRAM (e.g., up to \rev{38.5}$\times$ \omii{\omii{speedup with} 32GB of host DRAM}). This is because \popt's performance \revh{is} hindered by \omii{the host DRAM} capacity, while \proposal does not rely on large \omii{host} DRAM. Second, \aopt and {\aopt}+\cmashopt\ \revh{are not} affected by the small DRAM (except for the 32-GB configuration) due to their streaming database accesses.
But regardless of DRAM size, they suffer from I/O overhead.
Third, with the 32-GB DRAM\omii{, which is smaller than the extracted query k-mers in Step 1 (\sect{\ref{sec:mech-stage1-kmer-extraction}})},
\msfull's speedup increases. \omii{This is because}~\proposal's bucketing (\sect{\ref{sec:mech-stage1-kmer-extraction}}) \omii{avoids unnecessary page swaps between the host DRAM and the SSD in this case}.
\bback{We conclude that \proposal enables fast and accurate analysis, \emph{without relying} on large DRAM or large SSD-external bandwidth}.

\begin{figure}[h]
\centering
 \includegraphics[width=0.9\linewidth]{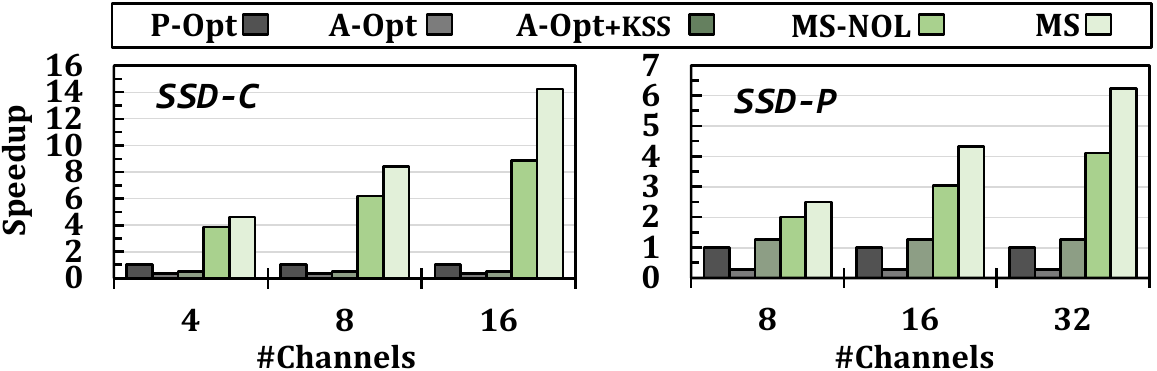}
\caption{\irev{Speedup with \omii{varying SSD internal bandwidth}}.}
\label{fig:internal-bw-swp}
\end{figure}

\head{\omii{Effect} of Internal Bandwidth} \revid{\label{rev:A4.3}A4.3}\fig{\ref{fig:internal-bw-swp}} shows the \omii{effect} of internal bandwidth (\omii{i.e.,} by varying the number of SSD channels) on \proposal with CAMI-M as a representative input. We observe that \proposals speedup increases as the internal bandwidth increases. 
On \ssdc (\ssdp), \proposal leads to 12.3--41.8x (8.6--21.6x) speedup over \aopt. \omii{The increased speedup is \omiii{due to the improved} performance of \proposals ISP steps as the internal bandwidth \omiii{increases}}.

\newcommand\poptp{\textsf{P-Opt\_P}\xspace}
\newcommand\aoptp{\textsf{A-Opt\_P}\xspace}
\newcommand\poptc{\textsf{P-Opt\_C}\xspace}
\newcommand\aoptc{\textsf{A-Opt\_C}\xspace}
\newcommand\msc{\textsf{MS\_C}\xspace}

\begin{figure}[t]
\centering
 \includegraphics[width=0.9\linewidth]{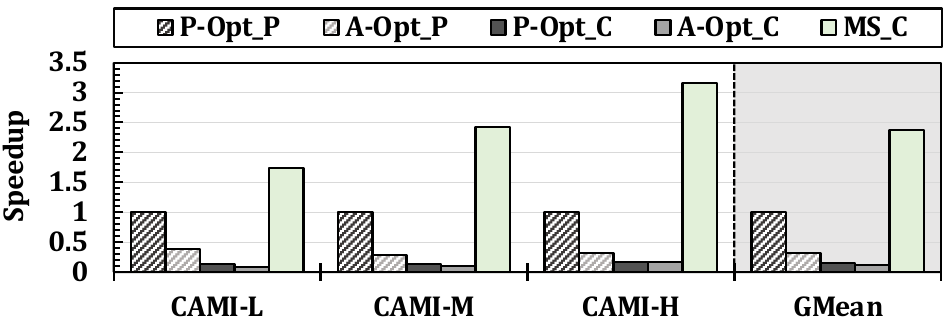}
\caption{Speedup of \proposal on a cost-optimized system over baselines on \omiii{cost-/}performance-optimized systems.}
\label{fig:ms-ce}
\end{figure}

\head{Impact on System Cost Efficiency} 
\proposal increases system cost-efficiency because \inum{i}~it analyzes large amounts of data inside storage and removes a large part of the analysis burden from other parts of the system, and \inum{ii}~\hhl{it} does not rely on either high-bandwidth host-SSD interfaces or large DRAM.
\fig{\ref{fig:ms-ce}}  compares \proposal on a cost-optimized system with \ssdc and 64-GB \omii{host} DRAM (\msc) to baselines \omiii{1) on the same system (\poptc and \aoptc) and 2) on} a performance-optimized system with  \ssdp and 1-TB \omii{host} DRAM (\poptp and \aoptp).\footnote{\omiii{For the performance-optimized system, we calculate the cost of 1TB DRAM to be roughly 7080 USD (8$\times$ 128GB modules~\cite{samsung128GBDDR4}) and the cost of \ssdp to be roughly 875 USD. For the performance-optimized system, we calculate the cost of 64GB DRAM to be roughly 312 USD (8$\times$ 8GB modules~\cite{samsung8GBDDR4}, assuming the same number of memory channels as the performance-optimized system) and the cost of \ssdc to be roughly 346 USD. Note that the cost of the \emph{total storage system} depends not only on the price of each SSD but also on the available interconnection slots in the systems, as systems typically have fewer PCIe slots (needed for \ssdp) than SATA slots (needed for \ssdc).}} \omiii{We make two key observations. First, \proposal on the cost-optimized system outperforms the baselines even when they run on the performance-optimized system.} \msc provides 2.4$\times$ and 7.2$\times$ average speedup compared to \poptp and \aoptp, respectively. Note that \msc provides the same accuracy as \aoptp and significantly higher accuracy than \poptp. 
\omiii{Second, baselines on the cost-optimized system experience significantly worse performance compared to when they run on the performance-optimized system. \poptc leads to 6.8$\times$ (7.7$\times$) average (maximum) slowdown over \poptp, and \aoptc leads to 2.8$\times$ (4.2$\times$) average (maximum) slowdown over \aoptp}. 
We conclude that \omiii{\proposal improves system cost-efficiency, while providing high performance and accuracy}.
This is critical \omii{to} both increasing the system cost-efficiency and enabling portable analysis, which is increasing\omii{ly} importan\omii{t due to} the advances of compact portable sequencers~\cite{minion21,jain2016oxford,cali2017nanopore} for on-site metagenomics\omii{~\cite{pomerantz2018real, chiang2019from,mutlu2023accelerating,alser2022molecules}}.

\begin{figure}[h]
\centering
\vspace{0.2em}
 \includegraphics[width=\linewidth]{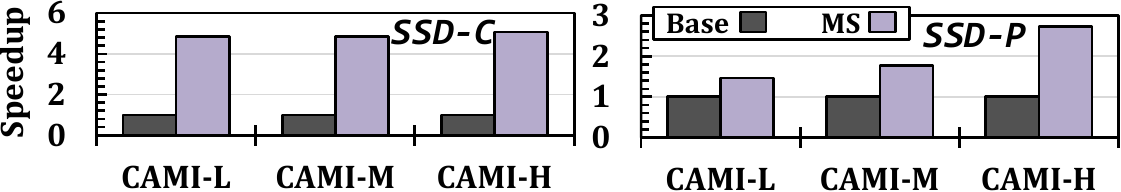}
\caption{\hm{Speedup over} a PIM-accelerated baseline~\cite{wu2021sieve}.}
\label{fig:eval-pim}
\end{figure}

\head{Comparison to a PIM Accelerator} \fig{\ref{fig:eval-pim}} compares \proposal to a PIM-accelerated baseline. We evaluate Kraken2's \emph{end-to-end} performance \bback{(i.e., including the I/O accesses to load data to the PIM accelerator, k-mer matching, sample classification, and other computation~\cite{wood2019improved})},
\hm{performing} k-mer matching \hm{on} a state-of-the-art
PIM system, Sieve~\cite{wu2021sieve}.\footnote{We do not use PIM for Metalign's k-mer matching as \omii{k-mer matching in Metalign} is bottlenecked only by I/O bandwidth, not main memory, due to its streaming accesses.} 
\proposal achieves 4.8-5.1$\times$ (1.5-2.7$\times$) speedup on \ssdc (\ssdp) \omii{over the PIM-accelerated baseline} while \omii{providing} significantly higher accuracy \omii{(4.8× higher F1 scores and 13\% lower L1 norm error)}.\footnote{A larger database for Kraken2 to encode richer information can increase accuracy \omii{for the PIM-accelerated baseline},
but with even large\hhl{r} I/O overhead.}

\subsection{Abundance Estimation Analysis}
\label{sec:eval-abundance}

\newcommand\msnidx{\textsf{MS-NIdx}\xspace}

\begin{figure}[h]
\centering
 \includegraphics[width=\linewidth]{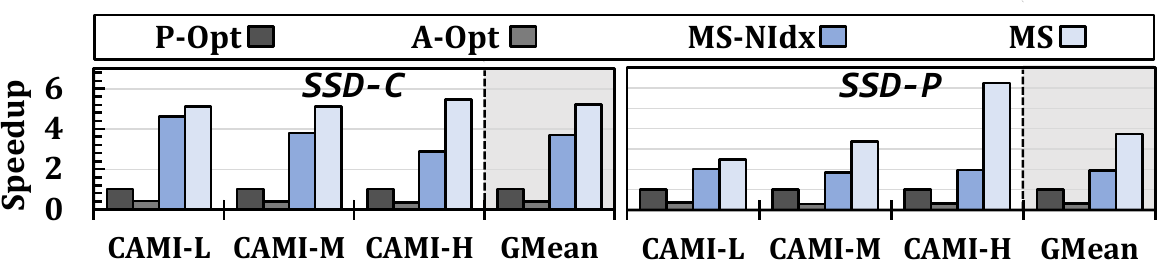}
\caption{\hm{Speedup} for abundance estimation.}
\vspace{0.3em}
\label{fig:eval-abundance}
\end{figure}

\omii{We evaluate abundance estimation with four configurations:
1)~\popt,
2)~\aopt, 
3)~\msnidx: a \proposal implementation that does not leverage {\proposal}'s third step for generating a unified reference index (\sect{\ref{sec:mech-stage1}}), and instead uses Minimap2~\cite{li2018minimap2} for index generation,
and 4)~\msfull: {\proposal}'s full implementation. 
\fig{\ref{fig:eval-abundance}} shows speedups over \popt}.
We make two key observations. 
First, 
\omii{\proposals full implementation} leads to significant speedup \omii{compared to both performance- and accuracy-optimized baselines. \msfull provides}
5.1--5.5$\times$ (2.5--3.7$\times$) speedup on \ssdc (\ssdp) compared to \popt, and 12.0--15.3$\times$ (6.5--20.8$\times$) speedup compared to \popt. Second, \proposal's full implementation achieves 65\% higher average speedup compared to \msnidx~\omii{due to \proposals efficient index generation}.

\subsection{Multi-Sample \omiii{Use Case}}
\label{sec:eval-multi-sample}

\newcommand\mspipe{\textsf{MS-\omiii{SW}}\xspace}

\begin{figure}[h]
\centering
 \includegraphics[width=0.9\linewidth]{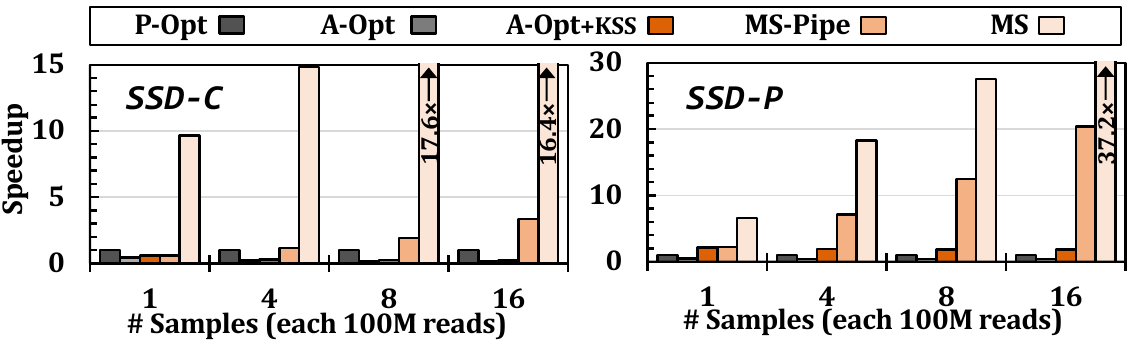}
\caption{\new{Speedup for multi-sample analysis.}}
\label{fig:multi-sample}
\end{figure}

\fig{\ref{fig:multi-sample}} shows speedup for \omiii{the} multi-sample use case \omiii{in which multiple samples need to access the same database} 
(\sect{\ref{sec:mech-multi-sample}}). 
\omiii{We consider} 256-GB host DRAM in which we can buffer k-mers \hhl{from} 1--16 samples.
We show the performance of \omiii{\proposals} multi-sample pipelined optimization (as described in \sect{\ref{sec:mech-multi-sample}}) in software (\mspipe) and in the full \proposal design (\msfull). In all configurations that require sorting \bback{(all except P-Opt)},
we use a state-of-the-art sorting accelerator~\cite{qiao2022topsort}.\footnote{\new{
We use the sorting throughput reported by the original paper~\cite{qiao2022topsort} and model the \omiii{data movement} time \omiii{between the sorting} accelerator \omiii{and other stages of \proposals pipeline}.}} 
First, \msfull achieves large speedup\omiii{s} of up to 37.2$\times$/100.2$\times$ \omiii{over} \popt/\aopt. Second, \mspipe leads to up to 20.5$\times$ (52.0$\times$) speedup \omiii{over} \aopt on \ssdc (\ssdp), and the speedup grows with the number of samples. 
\bback{We conclude that \omiii{\proposals pipeline optimization for the multi-sample use case in both software and hardware leads to large speedups over the baseline tools, and the hardware configuration leads to larger speedups compared to the software configuration by additionally leveraging ISP}}.

\subsection{Area and Power}
\label{sec:eval-area}

\joel{Table~\ref{tab:metastore_area_energy} shows the area and power consumption of \proposal's \omii{hardware accelerator} units \new{at 300 MHz}.
While the\omii{se units} could be designed to operate at a higher frequency, their throughput is already sufficient since \proposal is bottlenecked by NAND flash read \omii{throughput}}.
\proposals hardware \omii{accelerator} area and power requirements are \omii{small}: only 0.04~mm$^2$ and 7.658~mW at 65~nm.
\irev{\omii{The accelerator} can be place-and-routed in a small space with 0.25mm $\times$ 0.25mm dimensions (0.0625 mm$^2$)}.
The area overhead of \omii{the accelerator} is  0.011 mm$^2$ at 32 nm,\footnote{We scale area to lower technology nodes using the methodology in~\cite{stillmaker2017Scaling}.}
which is 1.7\% of the three 28-nm ARM Cortex R4 cores~\cite{cortexr4} in a SATA SSD controller~\cite{samsung860pro}. 
\bback{\omii{While both the accelerator and the cores in the SSD controller can execute \proposals ISP tasks, the accelerator is} 26.85$\times$ more power-efficient than the cores}.

\begin{table}[h]
\centering
\vspace{0.5em}
\caption{Area and power consumption of \proposal's logic.}
\label{tab:metastore_area_energy}
\resizebox{\columnwidth}{!}{%
\begin{tabular}{c|c|c|c}
\toprule
\textbf{Logic unit}                  & \textbf{\# of instances} & \textbf{Area [mm\textsuperscript{2}]} & \textbf{Power [mW]} \\ 
\midrule
\midrule
Intersect (120-bit)                 & 1 per channel              &    0.001361   &      0.284    \\
k-mer Registers (2$\times$ 120-bit)  & 1 per channel              &    0.002821   &      0.645    \\
Index Generator (64-bit)             & 1 per channel              &    0.000272   &      0.025    \\
Control Unit                         & 1 per SSD                  &    0.000188   &      0.026    \\\midrule
\textbf{Total for an 8-channel SSD}           & -                        & \textbf{0.04}    & \textbf{7.658} \\ \bottomrule
\end{tabular}
}
\end{table}

\icut{While\omn{We can remove this revision response since the topics are already covered in the previous sections.}\revid{\label{rev:B1.3}B1.3} \proposal introduces some changes to the system, its modifications are lightweight and we believe its large advantages justify the initial design effort. As discussed in \sect{\ref{sec:mechanism}}, \proposals lightweight ISP functionality can even run on the existing embedded SSD cores (while still requiring an ISP-specialized firmware). At this low cost, \proposal not only improves performance, but it also improves cost-efficiency (\fig{\ref{fig:ms-ce}}). This is critical because while sequencing cost has dropped significantly, computation cost has not been keeping up at the same rate\hhl{~\cite{berger2023navigating}}. Thus, \proposals contribution to high-performance, accurate, and cost-effective analysis paves the way for metagenomics’ broader adoption.}

\subsection{Energy}
\label{sec:eval-energy}

\omii{We demonstrate the energy consumption of different \mganalysis tools by obtaining the energy of the host processor, the host DRAM, the accelerators, the SSD's internal DRAM, host/SSD communications, and the SSD accesses}. 
\bback{For each tool, we calculate the energy \omii{consumption of each part of the system} based on its active/idle power and execution time. 
We observe that} 
\proposal provides significant energy benefits \omii{over other software and hardware baselines} by alleviating I/O overhead and
reducing the burden \omii{of \mganalysis} in the system \omii{(the host processor and DRAM)}.
\bback{\rev{A}cross \omii{our evaluated} SSDs and datasets}, 
\proposal leads to 5.4$\times$ (9.8$\times$), 15.2$\times$ (25.7$\times$), \hm{and} 1.9$\times$ (3.5$\times$) average (maximum) energy reduction compared to \popt, \aopt, and the PIM-accelerated \popt \bback{~when finding species present in the sample}. \omiii{By eliminating the need to move the large databases outside the SSD, \proposal leads to I/O data movement reduction of 71.7$\times$ over \aopt and 30.1$\times$ over \popt and the PIM-accelerated \popt.}

\section{Related Work}
\label{sec:related}

To our knowledge, \proposal is
the \emph{first} in-storage processing (ISP) system designed to significantly reduce the data movement overhead of \omf{the} end-to-end metagenomic analysis \omf{pipeline}.
By addressing the challenges of leveraging ISP for metagenomics, \proposal fundamentally alleviates its data movement overhead from the storage system \omii{via its efficient and cooperative pipeline between the host and the SSD}.

\head{Software Optimization of Metagenomics}
Several tools (e.g.,\omii{~\cite{lapierre2020metalign,song2024centrifuger,koslicki2016metapalette,Marcelino2020,piro2016dudes,piro2020ganon,pockrandt2022metagenomic,wood2014kraken}}) use comprehensive databases for high accuracy, but usually incur significant computational and I/O costs. 
Some tools (e.g.,\omii{~\cite{kim2016centrifuge,wood2019improved,muller2017metacache,song2024centrifuger,Dilthey2019,Fan2021}}) apply sampling to reduce database size, but at the cost of accuracy loss.

\head{Hardware Acceleration of Metagenomics}~\omii{S}everal works us\omii{e} GPUs \omii{(e.g.,}~\cite{jia2011metabing,kobus2021metacache,wang2023gpmeta,kobus2017accelerating,Su2012,su2013gpumetastorms,Yano2014}\omii{)},  FPGAs \omiii{(e.g.,~\cite{saavedra2020mining,zhang2023genomix,cervi2022metagenomic})}, and PIM \omii{(e.g.,}~\cite{wu2021sieve,shahroodi2022krakenonmem,shahroodi2022demeter,dashcam23micro,hanhan2022edam,zou2022biohd}\omii{)} to accelerate metagenomics by alleviating its computation or main memory overheads. 
The\omii{se works} do not \omii{reduce} I/O overhead\omii{s}, whose impact on 
\bback{end-to-end} 
performance becomes even larger when other bottlenecks are alleviated.
\irev{Some\revid{\label{rev:E3}E3} works~\cite{dunn2021,shih2023efficient} accelerate metagenomic analysis \omii{that use} raw genomic signals in targeted sequencing\omii{~\cite{kovaka2020targeted,Payne2021,Bao2021,AHMED2021102696}}. Targeted sequencing is not a focus of our work since this application looks for specific \omii{\emph{known}} targets in a sample, while we focus on cases where the contents of the sample are \omii{\emph{not known}} in advance and require looking up significantly larger databases. \icut{We believe analyzing ISP’s potentials for promising applications like targeted sequencing is an interesting future direction.}}

\irev{\head{Genome Sequence Analysis} Many works optimize different parts of \omii{the} genome analysis pipeline\omii{~\cite{mutlu2023accelerating,alser2022molecules}}. Several works (e.g.,\hhl{\omii{~\cite{huangfu2018radar,khatamifard2021genvom, cali2020genasm, gupta2019rapid,li2021pim,angizi2019aligns,zokaee2018aligner,turakhia2018darwin, fujiki2018genax, madhavan2014race,cheng2018bitmapper2,houtgast2018hardware,houtgast2017efficient, zeni2020logan,ahmed2019gasal2,nishimura2017accelerating,de2016cudalign,liu2015gswabe,liu2013cudasw++,liu2009cudasw++,liu2010cudasw++,wilton2015arioc,goyal2017ultra,chen2016spark,chen2014accelerating,chen2021high,fujiki2020seedex, banerjee2018asap,fei2018fpgasw,waidyasooriya2015hardware,chen2015novel,rucci2018swifold,haghi2021fpga,li2021pipebsw,ham2020genesis,ham2021accelerating,wu2019fpga}}}) accelerate read mapping, a commonly-used operation in genomics. As shown in \sect{\ref{sec:eval-abundance}}, \proposal can \omii{be} seamlessly integrate\omii{d} with different mappers. Some works (e.g.,\hhl{~\cite{saavedra2020mining,markus2020benchmarking,subramaniyan2021accelerated,fujiki2020seedex}}) optimize key primitives such as seeding. While these techniques have the potential to provide several benefits,
their adoption in metagenomics requires efficiently dealing with significantly larger and more complex indexes than the ones used in traditional genomics. Therefore, we hope that optimizations introduced in our work \omiii{can} facilitate the future adoption of these \omii{seeding techniques} in large-scale metagenomics}.

\head{In-Storage Processing}
Several works propose ISP as accelerators for different applications\omii{~\cite{liang2019cognitive,kim2020reducing,lim2021lsm,li2021glist,wang2016ssd,lee2020neuromorphic,kang2021s,han2021flash,wang2022memcore,wang2018three,han2019novel,choi2020flash,mailthody2019deepstore,pei2019registor,jun2018grafboost, do2013query, seshadri2014willow,kim2016storage, riedel2001active,riedel1998active,lee2022smartsage,jeong2019react, jun2016storage,li2023ecssd,wang2024beacongnn,jang2024smart}} (e.g., in machine learning\omii{~\cite{li2023ecssd,liang2019ins,lee2022smartsage,wang2024beacongnn,jang2024smart}}, pattern processing and read mapping~\cite{jun2016storage,mansouri2022genstore}, k-mer counting~\cite{abakus23taco}, \omii{and} graph analytics~\cite{jun2018grafboost}). Several \omii{works propose ISP in the form of}
general-purpose \omii{processing, inside the storage device~\cite{gu2016biscuit, kang2013enabling, wang2019project,acharya1998active,keeton1998case,riedel1998active,riedel2001active,merrikh2017high,tiwari2013active,tiwari2012reducing,boboila2012active,bae2013intelligent,torabzadehkashi2018compstor,kang2021iceclave,zou2022assasin}}, bulk-bitwise operations using flash~\cite{gao2021parabit,park2022flash},  \omii{SSDs closely integrated} with FPGAs~\cite{jun2015bluedbm, jun2016bluedbm, torabzadehkashi2019catalina, lee2020smartssd, ajdari2019cidr, koo2017summarizer}, or \omii{with} GPUs~\cite{cho2013xsd}. 
None of these works target the end-to-end metagenomic analysis. 
\irev{\revid{\label{rev:A6}A6}\proposal has two key differences from prior ISP approaches for genomics (e.g., pattern processing\cite{jun2016storage}, read mapping~\cite{mansouri2022genstore}, k-mer counting~\cite{abakus23taco}). First, \proposal is a cooperative ISP system for end-to-end metagenomic analysis that orchestrates processing both inside and outside the storage system, while other approaches focus on a specific task inside the storage system \bback{(e.g., pattern matching, read mapping filters, k-mer counting)}.
Second, while a part of \proposals pipeline (Part 1 of Step 2) performs the same functionality (i.e., sequence matching) as prior works, \proposal introduces new optimizations due to its unique requirements \omii{for cooperative ISP between the SSD and the host. As shown in \sect{\ref{sec:mech-stage2-1}},} these requirements stress the SSD's limited internal DRAM bandwidth as \proposal must 1) handle data from \emph{both} the host and SSD channels, and 2) share intermediate data across ISP stages efficiently}.

\section{Conclusion}
\label{sec:conclusion}

We introduce \proposal, the \emph{first} \omii{in-storage processing system} designed to significantly reduce the data movement overhead of end-to-end metagenomic analysis.  To enable efficient \omii{in-storage processing} for metagenomics, we propose new 1) task partitioning,
\omiii{2})~data/computation flow coordination,
\omiii{3})~storage-aware algorithms,
\omiii{4})~data mapping, 
and \omiii{5)~lightweight in-storage accelerators}.  \omii{We demonstrate that \proposal greatly improves performance, energy consumption, \omiii{and system} cost efficiency at low area and power costs}.

\omi{\section*{Acknowledgments}

We thank the anonymous reviewers of MICRO 2023, HPCA 2024, and ISCA 2024 for feedback. We thank the SAFARI group members for feedback and the stimulating intellectual environment. We acknowledge the generous gifts \omcviii{and support} provided by our industrial partners, \omii{including Google, Huawei, Intel, Microsoft, and VMware. This research was partially supported by \omiii{European Union’s Horizon Programme for research and innovation under Grant Agreement No. 101047160 (project BioPIM)}, the Swiss National Science Foundation (SNSF), Semiconductor Research Corporation (SRC), the ETH Future Computing Laboratory (EFCL), \omiii{and the AI Chip Center for Emerging Smart Systems Limited (ACCESS)}.}}

\bibliographystyle{unsrt}
\bibliography{refs}

\end{document}